\documentclass[letterpaper,11pt]{article}
\pdfoutput=1 
\usepackage[utf8]{inputenc}
\usepackage{jheppub} 
\usepackage{amsmath}
\usepackage{bbold}
\usepackage{xcolor}
\usepackage{graphicx}
\usepackage{hyperref}
\usepackage{mathtools}
\usepackage{jheppub}
\usepackage{amsfonts}
\usepackage{amssymb}
\usepackage{amsthm}
\usepackage{physics}
\usepackage{caption}
\usepackage{comment}
\usepackage{mathrsfs}
\usepackage{lscape}
\usepackage{amsmath,amssymb,amsfonts,leftindex,comment}
\usepackage{empheq}
\usepackage{subcaption}
\usepackage{amsthm}
\usepackage{mdframed}
\usepackage{enumitem}
\usepackage{orcidlink}
\usepackage{tabularx}
\theoremstyle{remark}

\usepackage{xcolor}

\title{On Unitarity of Bespoke Amplitudes}

\author[a,\orcidlink{0000-0001-7624-4421}]{Rishabh Bhardwaj}
\author[a,b,c, \orcidlink{0009-0008-2506-3207}]{Marcus Spradlin,}
\author[a,b, \orcidlink{0009-0005-6084-2466}]{Anastasia Volovich,}
\author[a, \orcidlink{0000-0003-3781-6153}]{He-Chen Weng}
\affiliation[a]{Department of Physics, Brown University,\\
	182 Hope Street, Providence, RI 02912, U.S.A.}
\affiliation[b]{Department of Physics, Harvard University,\\
        17 Oxford Street, Cambridge, MA 02138, U.S.A.}
\affiliation[c]{Brown Theoretical Physics Center, Brown University,\\
	340 Brook Street, Providence, RI 02912, U.S.A.}

\emailAdd{rishabh\_bhardwaj@brown.edu}
\emailAdd{marcus\_spradlin@brown.edu}
\emailAdd{anastasia\_volovich@brown.edu}
\emailAdd{he-chen\_weng@brown.edu}

\abstract{
We use partial wave unitarity to constrain various bespoke four-point amplitudes. We start by constructing bespoke generalizations of the type I superstring amplitude, which we show satisfy dual resonance and have suitable high-energy limits. By analyzing the behavior of partial wave coefficients for highly massive states, we strictly rule out all bespoke amplitudes with asymptotically non-linear Regge trajectories and place constraints on the first few non-trivial parameters in asymptotically linear cases. Finally, we argue that while a large class of unitary bespoke amplitudes fails to satisfy Regge Sum Rules, there exists a smaller sub-class with a vanishing mass gap that is superpolynomially bounded. 
}

\begin{document} 

\maketitle

\section{Introduction}

The primary goal of the bootstrap program of the 1960s was to derive scattering amplitudes from first principles such as Lorentz invariance, locality, and causality.  A triumph of this approach was the discovery of the Veneziano amplitude \cite
{Veneziano:1968yb}, which led to the discovery of string theory. Recently, there has been a resurgence of interest in bootstrap ideas, and a plethora of alternative candidate solutions to the bootstrap problem that attempt to generalize the Veneziano amplitude in various ways have been proposed  \cite{Coon:1969yw,Baker:1970vxk,Cheung:2022mkw,Geiser:2022exp,Cheung:2023adk,Cheung:2023uwn,Bhardwaj:2023eus,Haring:2023zwu,Arkani-Hamed:2023jwn} and studied \cite{Figueroa:2022onw,Wang:2024wcc,Rigatos:2023asb,Rigatos:2024beq,Eckner:2024ggx,Li:2023hce,Jepsen:2023sia,Bhardwaj:2022lbz,Chakravarty:2022vrp,Geiser:2022icl,Cheung:2024uhn}.
In particular, Cheung and Remmen introduced a new class of ``bespoke'' amplitudes that generalize the linear spectrum encoded in the Veneziano amplitude to allow for an arbitrarily customizable spectrum.  Specifically, the bespoke amplitude is defined by a Galois sum
\begin{equation}
    A_{\text{bespoke}}(s,t)=\sum_{\alpha,\beta} A_V(\nu_{\alpha}(s), \nu_{\beta}(t))\,,\label{eq:bespoke_amp_Galois_sum}
\end{equation}
over the roots $\{\nu_\alpha(\mu)\}$ of a  spectral curve of the form
\begin{equation}
    f(\mu,\nu)=P(\nu)-\mu Q(\nu) = 0 \label{eq:spectralcurve}
\end{equation}
for arbitrary polynomials $P(\nu)$ and $Q(\nu)$\footnote{We will always take the degrees of $P$ and $Q$ to satisfy
$|P|>|Q|$ to avoid having an accumulation point in the spectrum.},
where $A_V(s,t)$ is the Veneziano amplitude 
\begin{equation}
A_V(s,t)=\frac{\Gamma(-s) \Gamma(-t)}{\Gamma(-s-t)}\,.
\end{equation}
The Veneziano amplitude corresponds to the choice $P(\nu) = \nu$ and $Q(\nu) = 1$.
Bespoke amplitudes share the dual resonance property of the Veneziano amplitude,
but have finite high energy behavior as opposed to the exponentially soft behavior of the latter.

In this paper, we use the formalism developed in \cite{Arkani-Hamed:2022gsa} to study 
partial wave unitarity of bespoke amplitudes.
We rule out all bespoke amplitudes with asymptotically non-linear Regge trajectories. For models with asymptotically linear Regge trajectories, we constrain the first few coefficients in their spectral curves. Finally, we discuss the incompatibility of bespoke amplitudes with the Regge Sum Rules \cite{Haring:2023zwu}, which impose superpolynomial boundedness at high energies, which shows that most bespoke amplitudes are not string-like. However, we find that there is a small sub-class of bespoke models with a vanishing mass gap showcasing superpolynomial boundedness. 

The paper is organized as follows. In section \ref{sec:generalized_bespoke}, we generalize bespoke amplitudes to type I superstring theory amplitudes, which are dual resonant and have consistent high energy behavior.  In section \ref{sec:unitarity}  we study the unitarity of these amplitudes by computing the coefficients in their partial wave expansions.  In section \ref{sec:RSR_bespoke} we discuss
the consistency of these models with Regge Sum Rules. Finally, we conclude with some open problems in section \ref{sec:discussion}.

\section{Generalized bespoke amplitudes}\label{sec:generalized_bespoke}

In this section, we propose a generalization of (\ref{eq:bespoke_amp_Galois_sum}) that allows for bespoke amplitudes built from the type I superstring amplitude as well as the Veneziano amplitude. Specifically, we define
\begin{equation}
    A_{\text{bespoke}}^a = \sum_{\alpha,\beta}A^{a}_V(\nu_{\alpha}(s),\nu_{\beta}(t))\label{eq:a_deformed_bespoke}\,,
\end{equation}
where $\alpha$ and $\beta$ label roots of the spectral curve (\ref{eq:spectralcurve}) and now, following the conventions of \cite{Arkani-Hamed:2022gsa}, we take
\begin{equation}
A^a_V(s,t) =- s^{a-1}\frac{\Gamma(1-2a-s)\Gamma(-a-t)}{\Gamma(1-s-t-3a)}\,.
\end{equation}
The special case $a=0$ corresponds to the type I string amplitude
\begin{align}
    A^{a=0}_V(s,t)=
\frac{\Gamma(-s)\Gamma(-t)}{\Gamma(1-s-t)}
\end{align}
while $a=1$ gives the (tachyonic) bosonic string amplitude. Specifically, we note that
\begin{align}
A^{a=1}_V(s,t) = - \frac{\Gamma(-s-1)\Gamma(-t-1)}{\Gamma(-s-t-2)} = - \frac{(1+s+t)(2+s+t)}{(1+s)(1+t)} A_V(s,t)
\end{align}
differs from the convention for $A_V(s,t)$ used in \cite{Cheung:2023uwn} by some prefactors of the Mandelstam variables that are qualitatively unimportant for our analysis. In the rest of this section, we check that (\ref{eq:a_deformed_bespoke}) satisfies dual resonance and we determine its high energy behavior in the Regge and hard scattering limits.

\subsection{Dual resonance}

We now check that the generalized bespoke amplitudes satisfy dual resonance. An amplitude satisfies dual resonance if the residue $R_n(t)$ of each pole in its $s$-channel expansion
\begin{align}
A(s,t) = \sum_{n=0}^\infty \frac{R_n(t)}{s - m_n^2}
\end{align}
is a polynomial in $t$ (and, by crossing symmetry, vice versa). For the bespoke amplitude defined in (\ref{eq:a_deformed_bespoke})
we compute the corresponding residue polynomials in (\ref{eq:residue_a_deformed_bespoke}).

We start by noting that the amplitude manifestly inherits the crossing symmetry from that of the $a$-deformed Veneziano amplitude. Further, the simple pole structure of \eqref{eq:a_deformed_bespoke} can be seen by starting from the contour integral representation
\begin{equation}
        A_{\text{bespoke}}^a(s,t) = \frac{1}{(2\pi i)^2}\oint\sum_{\alpha}\frac{d\sigma}{\sigma-\nu_{\alpha}(s)}\oint\sum_{\beta}\frac{d\tau}{\tau-\nu_{\beta}(t)}A_{V}^a(\sigma,\tau)\label{eq: a_deformed_bespoke_contour_def}~.
\end{equation}
The $a$-Veneziano amplitude $A^a_V(\sigma,\tau)$ has the integral representation \cite{Arkani-Hamed:2022gsa}
\begin{equation}
    A^a_V(s,t) = -s^{a-1}\int_{0}^1dz~z^{-s-2a}(1-z)^{-1-t-a} \label{eq:integral_rep_Veneziano}
\end{equation}
with simple poles at positive integer values i.e.~$\sigma,\tau = n-a$ for $n\in \mathbb{N}$, which implies that \eqref{eq: a_deformed_bespoke_contour_def} only has simple poles at $\nu_{\alpha}(t)=n-a$. 

Next, we must verify whether the amplitude has polynomial residues. This prohibits the amplitude from developing higher-order poles when both Mandelstam variables approach the same singularity. We will compute the residue at $\nu_{\alpha}(s)=n-a$ and demonstrate that it is a polynomial in the variable $t$. Let us start by plugging the representation
\begin{equation}
    A_{V}^a(\sigma,\tau) = \sum_{n=0}^{\infty}\frac{R_V^a(n-a,\tau)}{n-a-\sigma}
\end{equation}
for the $a$-Veneziano amplitude into \eqref{eq: a_deformed_bespoke_contour_def}, which leads to
\begin{equation}
    A_{\text{bespoke}}^a(s,t) = \sum_{n=0}^{\infty}\sum_{\alpha,\beta}\frac{R^a_V(n-a,\nu_{\beta}(t))}{n-a-\nu_{\alpha}(s)}~.
\end{equation}
From this, it is straightforward to obtain the residue
\begin{equation}
    R^a_{\text{bespoke}}(t) = \lim_{s\to \mu(n-a)}(\mu(n-a)-s)A^a_{\text{bespoke}}(s,t)
\end{equation}
for which we first notice 
\begin{align}
    \sum_{\alpha} \frac{1}{n-a-\nu_{\alpha}(s)} &= \partial_n\log\left(\prod_{\alpha} (n-a-\nu_{\alpha(s)})\right)
    = \partial_n \log f(s,n-a)\,.
\end{align}
Plugging in the spectral curve (\ref{eq:spectralcurve}) gives
\begin{align}
    \sum_{\alpha} \frac{1}{n-a-\nu_{\alpha}(s)} 
    & = \frac{\mu'(n-a)}{\mu(n-a)-s}+\frac{Q'(n-a)}{Q(n-a)}
\end{align}
which leads to
\begin{equation}
    R^a_{\text{bespoke}}(t) = \mu'(n-a)\sum_{\beta}R^a_{V}(n-a,\nu_{\beta}(t))~.\label{eq:Res_a_deformed_bespoke}
\end{equation}
Note that even though dual resonance of the $a$-Veneziano amplitude implies that $R^a_{V}(n-a,\nu_{\beta}(t))$ must be a polynomial in $\nu_{\beta}(t)$, it need not be a polynomial in $t$ since $\nu_{\beta}(t)$ is in general an algebraic function of $t$. To prove that \eqref{eq:Res_a_deformed_bespoke} is a polynomial, note the identity
\begin{align*}
    R^a_{V}(n,x) 
    &= \frac{(-n)^{a-1}}{(n+2a-1)!}\frac{\Gamma(3a+n+x)}{\Gamma(1+a+x)} = \frac{(-n)^a}{(n+2a-1)!}(x+a+1)^{(2a+n-1)}
\end{align*}
which follows from the Gamma function representation of the $a$-Veneziano amplitude and Euler's reflection identity. Further utilizing Newton's identity for the rising factorial $x^{(a)}$ we arrive at
\begin{equation}
    R^a_{V}(n-a,x) = \frac{(a-n)^{a-1}}{(n+a-1)!}\sum^{n+a-1}_{k=0}\begin{bmatrix}
           n+a-1 \\
           k+1 
         \end{bmatrix}(x+a+1)^k
\end{equation}
where the quantity in square brackets is the Stirling number of the first kind. Therefore, plugging the above back into \eqref{eq:Res_a_deformed_bespoke}, we finally have that the residue of \eqref{eq:a_deformed_bespoke} is given as
\begin{equation}
    R^a_{\text{bespoke}}(t) = \frac{\mu'(n-a)(a-n)^{a-1}}{(n+a-1)!}\sum^{n+a-1}_{k=0}\begin{bmatrix}
           n+a-1 \\
           k+1 
         \end{bmatrix}d_k(t)\label{eq:residue_a_deformed_bespoke}
\end{equation}
where we have defined
\begin{equation}
    d^a_{k}(t) = \sum_{\beta}(\nu_{\beta}(t)+a+1)^k~.
\end{equation}
From here on, the proof follows identically from the one mentioned in \cite{Cheung:2023uwn} for the $a=1$ case, where the authors utilize key results from Galois theory on elementary symmetric polynomials and Newton's theorem to show that $d_k(t)$ are polynomials in $t$ for all $k$.\footnote{Note that $d^a_k(\nu_{\alpha}(t))$ is related to the symmetric polynomial $d_k(\nu_{\alpha}(t))$ defined in \cite{Cheung:2023uwn} up to a simple shift $\nu_{\beta}(t)\to \nu_{\beta}(t)+a+1$. This will not affect the re-utilization of the results in \cite{Cheung:2023uwn} for our proof. To see this, note that 
\begin{equation}
    d^a_k(t) = \sum_{m=0}^{k}\binom{k}{m}(a+1)^{k-m}d_{m}(t);
\end{equation}
thus if $d_{m}(t)$ is a polynomial in $t$ for any $m \in \mathbb{N}$ then so is $d^a_k(t)$.} This concludes the proof that \eqref{eq:a_deformed_bespoke} is dual resonant.

\subsection{High energy asymptotics}

\subsubsection{Regge limit}

    In this section, we determine the behavior of the generalized bespoke amplitude \eqref{eq:a_deformed_bespoke} in the Regge limit $s \to \infty$ with $t$ held fixed.

   In this limit, the zeros of the spectral function $f(s,\sigma) = P(\sigma)-sQ(\sigma) = 0$ will be dominated by the roots of the $Q(\sigma)$ polynomial. The only exception is when $|Q| = |P|-1\equiv h-1$ and  $\sigma = s$, in which case the degrees of the two polynomials are comparable, and they could cancel each other out. In the end, we are only left with the following two possibilities for the roots $\nu_{\alpha}(s)$
    \begin{equation}
        \lim_{s\to \infty} \nu_{\alpha}(s) = \begin{cases}
            s\,,\qquad \alpha = 0\\
            \nu_{\alpha}^{(Q)}\,,~~ \alpha=1,\dots,h-1
        \end{cases}\label{eq:roots_regge_limit}
     \end{equation}
     with the order of roots being chosen without loss of generality. Then, in this limit, one gets
     \begin{align}
        A^a_{\infty}(t) &\equiv \lim_{s\to \infty} A_{\text{bespoke}}^a(s,t) 
     \sim \sum_{\beta} \left(s^{\nu_{\beta}(t)+2a-1}+\sum_{\alpha \neq 0}(\nu_{\alpha}^{(Q)})^{\nu_{\beta}(t)+2a-1}\right)
     \end{align}
     where we have utilized the Regge limit of the $a$-Veneziano amplitude $A^a_V(s,t) \sim s^{t+2a-1}$. The above Regge limit only exists when the bound
     \begin{equation}
         \text{Re}(\nu_{\beta}(t))< 1-2a\,,\qquad \forall~\beta \label{eq:regge_scattering_constraint}
     \end{equation}
     is satisfied, in which case the amplitude has the non-zero boundary term at infinity
     \begin{equation}
         A^a_{\infty}(t) \sim \sum_{\beta,\alpha \neq 0}(\nu_{\alpha}^{(Q)})^{\nu_{\beta}(t)+2a-1}\,.\label{eq:a_deformed_bespoke_bdry_term}
     \end{equation}
The presence of boundary terms seems to be in direct contradiction with the superpolynomial boundedness that string amplitudes are known to possess. Such terms modify the dispersion relations which allow one to rewrite a dual resonant four-point amplitude as an infinite sum over massive states, as a residue arising from a pole at infinity.  Such asymptotic behavior is consistent with locality of quantum field theory. We elaborate more on this in section \ref{sec:RSR_bespoke}. One can make choices of the spectral curve such that the above boundary term vanishes at least when \eqref{eq:regge_scattering_constraint} holds. Some obvious ones are $Q(\nu) = \nu^{|Q|}$ such that $\nu_{\alpha}^{(Q)}=0$ for all $\alpha$ or simply $|Q|=0$. The former corresponds to the post-Regge bespoke models with a vanishing mass gap, which we will introduce in section \ref{sec:post_linear_regge_unitarity}.

     \subsubsection{Hard scattering limit} 
     
 In this section, we determine the behavior of the generalized bespoke amplitude \eqref{eq:a_deformed_bespoke} in the hard scattering limit at a fixed angle, which corresponds to $s,t \to \infty$ with $s/t$ fixed.

     In this limit, the generalized Veneziano amplitude is
     \begin{equation}
         A^{a}_{\text{hard},V}(s,t) \sim \exp\left(B_{a}(s,t)\right)
     \end{equation}
     where
     \begin{align}
         B_{a}(s,t) = &~(2a+s-1)\log(2a+s-1)-(a+t)\log(a+t)\nonumber\\&+(3a+s+t-1)\log(3a+s+t-1)+(a-1)\log s\,.
     \end{align}
     Therefore in this limit \eqref{eq:a_deformed_bespoke} takes the form 
     \begin{align}
         &A^{a}_{\text{hard,bespoke}} = \lim_{s,t \to \infty} A^{a}_{\text{bespoke}}(s,t)\nonumber\\
         &= \lim_{s,t \to \infty} A^{a}_{\text{hard},V}(s,t)+\lim_{s\to \infty} \sum_{\alpha \neq 0} A^a_V(s,\nu_{\alpha}^{(Q)})+\lim_{t\to \infty}\sum_{\beta \neq 0} A^a_V(\nu_{\beta}^{(Q)},t)+\sum_{\alpha,\beta \neq 0} A^a_V(\nu_{\alpha}^{(Q)},\nu_{\beta}^{(Q)})\nonumber\\
         &\sim \exp\left[B_{a}(s,t)\right]+\sum_{\alpha \neq 0}\left(s^{2a-1+\nu_{\alpha}^{(Q)}}+t^{2a-1+\nu_{\alpha}^{(Q)}}\right)+\sum_{\alpha,\beta \neq 0} A^a_V(\nu_{\alpha}^{(Q)},\nu_{\beta}^{(Q)})
     \end{align}
     where in going from the first to the second line, we have used the expression for the roots of $Q$ in the Regge limit \eqref{eq:roots_regge_limit}. For the above limit to exist, the RHS of the expression should vanish.
    The first term vanishes when $t > -a$, and the second and third terms vanish when
     \begin{equation}
         \text{Re}(\nu_{\alpha}^{(Q)}) < 1-2a\,,\qquad\forall~\alpha\,, \label{eq:hard_scattering_constraint}
     \end{equation}
     which is an $a$-generalization of the result presented in \cite{Cheung:2023uwn}.
    In conclusion, when these conditions are satisfied, the generalized bespoke  in the hard scattering limit takes the form
     \begin{equation}
          A^{a}_{\text{hard,bespoke}} = \sum_{\alpha,\beta \neq 0} A^a_V(\nu_{\alpha}^{(Q)},\nu_{\beta}^{(Q)})\,.\label{eq:hard_a_deformed_bespoke}
     \end{equation}
     
     Here, we see the presence of a boundary term like we saw in the Regge limit \eqref{eq:a_deformed_bespoke_bdry_term}. As in the Regge limit case, there exist particular choices of the polynomial $Q$, which can make the above boundary term vanish. To see this, first notice that these choices could simply correspond to poles of the denominator of \eqref{eq:a_deformed_bespoke}. These zeros occur when $\nu^{(Q)}_{\alpha}+\nu^{(Q)}_{\beta} \in \mathbb{Z}_{\geq 1-3a}$ while $\nu^{(Q)}_{\alpha},\nu^{(Q)}_{\beta} \in \mathbb{C}/\mathbb{Z}_{\geq -a}$ for all $\alpha,\beta$. In general, this implies $\text{Im}(\nu^{(Q)}_{\alpha})=0$ for all $\alpha$ and $\text{Re}(\nu^{(Q)}_{\alpha}) = \frac{n_{\alpha}}{2}$ for some $n_{\alpha}\in \mathbb{Z}_{\geq {1-3a}}$, while making sure $n_{\alpha}+n_{\beta}\in \mathbb{Z}_{\geq {2-6a}}$ whenever $\alpha\neq \beta$. But the hard scattering constraint \eqref{eq:hard_scattering_constraint} implies $n_{\alpha}<2-4a$ for all $\alpha$. The two conditions imply that $n_{\alpha}$ only exists if $a=0$, for which $n_{\alpha}=1$ identically. This is possible only when $Q(\nu) = (\nu-\frac{1}{2})^{|Q|}$; these are once again the superstring post-Regge bespoke models with mass gap $-\frac{1}{2}$.

\section{Unitarity of generalized bespoke amplitudes}\label{sec:unitarity}

In this section, we study the unitarity of the generalized bespoke models. As a warm-up, we first consider in section \ref{sec:KK_unitarity_toy_example} the bespoke amplitude with an asymptotically Kaluza-Klein (KK) spectrum that is defined in (\ref{eq: Kaluza_Kline spectrum}). Next, in section \ref{sec:unitarity_non_linear_regge}, we rule out general bespoke amplitudes with asymptotically non-linear Regge trajectories. Finally, in section \ref{sec:unitarity_linear_regge}, we consider bespoke models with asymptotically linear Regge trajectories, with a particular focus on a certain class of models possessing a post-Regge expansion. The techniques and derivations in this section are largely inspired by and closely mirror the ones presented in \cite{Arkani-Hamed:2022gsa, Bhardwaj:2022lbz}.

\subsection{Warm-up example: 
generalized Kaluza-Klein amplitude }\label{sec:KK_unitarity_toy_example} 

The spectral curve of the KK theory is given by the quadratic polynomial
\begin{equation}\label{eq: Kaluza_Kline spectrum}
    f(\mu,\nu) = (\nu + \delta)^2 - \mu\,.
\end{equation}
The roots are $\nu_{\pm} = -\delta \pm \sqrt{\mu}$~, and the bespoke amplitude is 
\begin{align}\label{eq:Kaluza_Kline}
   A^a(s,t) &= A^a_V({-}\delta {+} \sqrt{s},{-}\delta {+} \sqrt{t}) + A^a_V({-}\delta {+} \sqrt{s},{-}\delta {-} \sqrt{t}) \nonumber \\
   &~~~+ A^a_V({-}\delta {-} \sqrt{s},{-}\delta + \sqrt{t}) {+} A^a_V({-}\delta {-} \sqrt{s},{-}\delta {-} \sqrt{t})\,.
\end{align}
The amplitude has poles at $s=(n+\delta)^2$ and the residues are analytic.  The amplitude in its worldsheet integral representation is given as 
\begin{align}\label{eq:kaluza-kline_world sheet_integral}
    &A^a(s,t)=-\int_{0}^{1}dz~\left((-\delta+\sqrt{s})^{a-1}z^{\delta-\sqrt{s}-2a}+(-1)^{a-1}(\delta+\sqrt{s})^{a-1}z^{\delta+\sqrt{s}-2a}\right)\nonumber\\
    &\qquad\qquad\qquad\quad\times\left((1-z)^{\delta-\sqrt{t}-1-a}+(1-z)^{\delta+\sqrt{t}-1-a}\right)
\end{align}
To conduct the partial wave unitarity analysis, we follow the line of approach developed in \cite{Arkani-Hamed:2022gsa,Bhardwaj:2022lbz}. Unlike there, it is not feasible to find a double contour representation for the partial wave coefficients of the above amplitude. However, this is achievable in a particular asymptotic limit in which an exchanged resonance of a given spin becomes infinitely massive. The corresponding representations for the coefficients that arise in this limit will allow us to calculate some exact expressions of them. 

Starting from the integral representation of the Veneziano amplitude \eqref{eq:integral_rep_Veneziano}, one can see that the $s=n-4a$ poles of the amplitude are generated from the singularity $z^{-s-1}$ around zero. Therefore, it is natural to expand the integrand around $z=0$ and evaluate the integral. Indeed, Taylor expanding the $t$ dependent part $(1-z)^{-t-1}$ around $z=0$ allows us to rewrite the integrand in the form of a series $\sum_m a_m(t)z^m$. Then after doing the $z$ integral the amplitude can be evaluated to be $A^a_V(s,t)=\sum_m\frac{a_m(t)}{-s+m}$. Thus, the residue at $s=n$ is $R^{a}_{V,n}(t)=a_n(t)$, which is the $n$-th derivative of $(1-z)^{-a-t-1}$ evaluated at zero and can be computed via the contour integral 
\begin{equation}
    R^{a}_{V,n}(t)=n^{a-1}\oint_{z=0}\frac{dz}{2\pi i z^{n+2a}}(1-z)^{-a-t-1}\,.
\end{equation}
Similarly, for the KK amplitude (\ref{eq:kaluza-kline_world sheet_integral}), the residue at $s=(n+\delta)^2$ can also be expressed as
\begin{equation}
    R^a_{\text{KK},n}(t)= J_{n,\delta}\oint_{z=0}\frac{dz}{2\pi i z^{n+2a}}\left((1-z)^{\delta-\sqrt{t}-1-a}+(1-z)^{\delta+\sqrt{t}-1-a}\right).
\end{equation}
Here $J_{n,\delta}{=}2(n{+}\delta)n^{a-1}$ is the Jacobian factor, which we will suppress for now and call back in the final answer. Introducing the variable $x=\cos\theta$ and using the relation $t=\frac{s+4a}{2}(x-1)=\frac{(n+\delta)^2+4a}{2}(x-1)$, we have
\begin{align}
    R^a_{\text{KK},n}(y)
    &= \oint_{u=0}\frac{du}{2\pi i }\frac{ 2e^{-u(-3a+\delta-n)}}{(e^u-1)^{n+2a}} \cos (u (n+\delta) y)~,
\end{align}
where we made the change of variables $y=i \sqrt{\left(1+\frac{4a}{(n+\delta)^2}\right)\frac{x-1}{2}}$ \footnote{Note that the quantity $\sqrt{\left(1+\frac{4a}{(n+\delta)^2}\right)\frac{x-1}{2}}$ is always zero or pure imaginary. Therefore, we introduce an additional imaginary unit in the definition of $y$ to ensure it is a real number.} and $z=1-e^{-u}$.
We would like to expand the residues in the basis of  Gegenbauer polynomials $G^{(D)}_j(x)$, which are orthogonal with respect to the measure $d\theta (\sin\theta)^{D-3}=dx (1-x^2)^{\frac{D-4}{2}}$. A particularly useful representation of $G^{(D)}_j(x)$ is given by the Rodrigues formula 
\begin{equation}
    G^{D}_j(x)=(-1)^{j}\alpha_{j,D}(1-x^2)^{-\delta_0}\partial_x^j (1-x^2)^{j+\delta_0},
\end{equation}
where $\alpha_{j,D} = \frac{(2\delta_0+1)_j}{2^j j! (\delta_0+1)_j}$ and $\delta_0 = \frac{D-4}{2}$. For any function $F(x)$ that can be expanded on the Gegenbauer polynomials $F(x)=\sum_j F_j G^{(D)}_j(x)$, the corresponding coefficients $F_j$ can be obtained via orthogonality as
\begin{equation}
F_j = n_{j,D} \int_{-1}^1 dx (1-x^2)^{\delta_0} F(x) G^{(D)}_j(x)\,,\, \qquad n_{j,D}=j!\frac{2^{2\delta_0}(j+\delta_0+\frac{1}{2})\Gamma(\delta_0+\frac{1}{2})^2}{\pi\Gamma(j+2\delta_0+1)}\,.
\end{equation}
Both constants, $\alpha_{j,D}$ and $n_{j,D}$, are manifestly positive, so we will set them aside for convenience and pick them up later. Applying the Rodrigues formula, we then get
\begin{align}\label{partial_wave_coefficient_projection}
    F_j&=(-1)^j\int^1_{-1}dxF(x)\partial^j_x(1-x^2)^{j+\delta_0}~,\nonumber\\
    &=-\int^0_{-1} dy'4y' (4y'^2)^{(j+\delta_0)}(1-y'^2)^{(j+\delta_0)}\left(\frac{-1}{4y'}\partial_{y'}\right)^j F(y')\,,
\end{align}
where in the second line, we perform integration by parts $j$ times and assume that $F(x)$ is regular at $x=-1,\,1$. Furthermore, we make an additional change of variable $y'=\frac{y}{\sqrt{1+\frac{4a}{(n+\delta)^2}}}=i\sqrt{\frac{x-1}{2}}$. From now on, we will drop the primes from our $y$ variables.
Next, we expand the residue on the basis of Gegenbauer polynomials
\begin{equation}
    R^a_{\text{KK},n}(x)=\sum_j B^{a(D)}_{n,j} G^{(D)}_{j}(x)\,,
\end{equation}
where the coefficients $B^{a(D)}_{n,j}$ are given by 
 \begin{align}
 &B^{a(D)}_{n,j} 
    = 2^{2(j+\delta_0+1)}\oint_{u=0}\frac{du }{2\pi i }\frac{e^{-u(-3a+\delta-n)}}{(e^u-1)^{n+2a}}\mathcal{H}^{a,\delta_0}_{n,j}(u,\delta)~,\nonumber\\
    &\text{and}~~     \mathcal{H}^{a,\delta_0}_{n,j}(u,\delta)
    \equiv 2\int^{1}_{0} dy\  y^{2(j+\delta_0)+1}(1-y^2)^{j+\delta_0}\left(\frac{-1}{4y}\partial_y\right)^j  \cos \left(u (n+\delta)\sqrt{1+\frac{4a}{(n+\delta)^2}} y\right)~,\label{eq:partial_wave_coeff_H_int_KK}
\end{align}
 where in the second line, we exchanged the integration order and made the change of variable $y \rightarrow -y$. To proceed we first analyse how the differential operator $\frac{-1}{y}\partial_y$ acts on $\cos(\alpha y)$, writing
\begin{equation}
    \left(\frac{-1}{y}\partial_y\right)^j  \cos (\alpha y)=\alpha^{2j} \sqrt{\frac{\pi}{2}} (\alpha y)^{\frac{1}{2}-j} J_{\frac{1}{2}(2j-1)}(\alpha y)
\end{equation}
in terms of half-integer Bessel functions. This will allow us to rewrite $\mathcal{H}^{a,\delta_0}_{n,j}$ as 
\begin{equation}\label{eq:H_integral_KK}
    \mathcal{H}^{a,\delta_0}_{n,j}(u,\delta) = 2^{1-2j}\alpha^{\frac{1}{2}+j}\sqrt{\frac{\pi}{2}}\int^{1}_{0} dy\  y^{2(j+\delta_0)+1}(1-y^2)^{j+\delta_0}   y^{\frac{1}{2}-j} J_{\frac{1}{2}(2j-1)}(\alpha y)\,,
\end{equation}
where $\alpha=i u\sqrt{(n+\delta)^2+4a}$~.
 We can then use the above result combined with \eqref{eq:partial_wave_coeff_H_int_KK} to calculate the partial wave coefficients for large $n$. The generating function \cite{238f62ea-f226-315b-a579-8f3544634def,watson1995treatise}
\begin{equation}
    \sum_{j=0}^{\infty} J_{j-\frac{1}{2}}(r)\frac{t^j}{j!} = \sqrt{\frac{2}{\pi r}}\cos \sqrt{r^2-2rt}\,,
\end{equation}
allows us to rewrite the partial wave coefficients as
\begin{align}
    B^{a(D)}_{n,j} 
    &= 4^{\delta_0+1} (((n{+}\delta)^2+4a))^{\frac{j}{2}}j!\oint_{u=0}\frac{du}{2\pi i}\frac{ e^{-u(-3a+\delta-n)}}{ (e^u-1)^{n+2a}} u^{j}{\int^1_0} dy ~y^{1+2\delta_0+j}\left(1{-}y^2\right)^{\delta_0+j} \nonumber\\
    &~~~\times{\oint_{t=0}} \frac{dt}{2\pi i t^{j+1}} \left(e^{\sqrt{(\alpha y)^2{-}2it \alpha y}}+e^{- \sqrt{( \alpha y)^2{-}2it \alpha y}}\right)\,.
\end{align}
 In the large $n$ limit, $|\alpha|$ approaches infinity, allowing us to expand $e^{\sqrt{\alpha^2y^2-2it \alpha y}} = e^{ \alpha y\sqrt{1-\frac{2it}{\alpha y}}}\sim e^{\alpha y-it}.$
Therefore, in this limit, we have
\begin{align}
    B^{a(D)}_{n,j}&\sim~4^{\delta_0+1} (((n{+}\delta)^2+4a))^{\frac{j}{2}}j!\oint_{u=0}\frac{du}{2\pi i}\frac{ e^{-u(-3a+\delta-n)}}{ (e^u-1)^{n+2a}} u^{j}{\oint_{t=0}} \frac{dt}{2\pi i t^{j+1}}\nonumber\\
    &~~~~\times{\int^1_0} dy ~y^{1+2\delta_0+j}\left(1{-}y^2\right)^{\delta_0+j}\left(e^{\left( \alpha y-it\right)}{+}e^{-\left(\alpha y-it\right)}\right)\nonumber\\
 \end{align} 
and upon doing the $y$ and $t$ integral we arrive at
 \begin{align}
 B^{a(D)}_{n,j}&\sim 4^{\delta_0+1}(-i)^j (((n{+}\delta)^2+4a))^{\frac{j}{2}}\nonumber\\
 &~~~\times\oint_{u=0}\frac{du}{2\pi i}\frac{ e^{u(3a+n-\delta)}}{ (e^u-1)^{n+2a}} u^{j}\partial_{\alpha}^{1+2\delta_0+j}\left(1{-}\partial_{\alpha}^2\right)^{\delta_0+j}\left(\frac{-2{+}e^{\alpha}{+}e^{-\alpha}}{\alpha}\right)\,.
\end{align}
Here we restrict to even $D$, so that $\delta_0$ is an integer. As we take the large $|\alpha|$ limit, we can drop the first term in $\frac{-2{+}e^{\alpha}{+}e^{-\alpha}}{\alpha}$, since depending on the sign of $\text{Im}(u)$ either of the two exponentials pieces always dominate in the expression. Then, applying the identity
\begin{equation}
   (1-\partial_{\alpha}^2)^J\left(\frac{e^{\alpha}}{\alpha}\right)=J!e^{-\alpha}\partial_{\alpha}^J\left(\frac{e^{2\alpha}}{\alpha^{J+1}}\right)\label{eq:diagonalisation_identity}
\end{equation}
proven in \cite{Arkani-Hamed:2022gsa} and further using Cauchy's integral formula $\partial^{J}_uf(u) = (-1)^JJ!\oint_{v=0}\frac{dv}{2\pi i}\frac{f(u-v)}{v^{J+1}}$ for any smooth function $f$ we get the following triple-contour representation for the partial wave coefficients in the large $n$ and fixed $j$ limit:
\begin{align}
    &B^{a(D)}_{n,j}\sim  \mathcal{C}^{D}_{n,j}(\delta)\oint_{u=0}\frac{du}{2\pi i}\frac{ e^{-u(-3a+\delta-n)}}{ (e^u-1)^{n+2a}} u^{j}\oint_{w=0}\frac{dw}{2\pi i ~w^{2+2\delta_0+j}} \oint_{v=0}\frac{dv}{2\pi i ~v^{\delta_0+j+1}}\frac{\cosh{(\alpha-w-2v)}}{(\alpha{-}w{-}v)^{1+\delta_0+j}}\,,\nonumber
    \end{align}
where $\mathcal{C}^{D}_{n,j}(\delta)=2^{2\delta_0+3}(-1)^{1+\delta_0+\frac{3j}{2}}n^{j}((\delta_0{+}j)!)^2(1{+}2\delta_0{+}j).$
Again since $|\alpha|~{\gg}~w\ \text{and}\ v$, we can drop the $w$ and $v$ terms in the denominator $\alpha{-}w{-}v$ of the $v$ contour integral. Further we can make the approximation $((n+\delta)^2+4a)^{\frac{j}{2}} \sim n^j$ in this limit. Then after doing the $v$ and $w$ integrals and plugging back in the value of $\alpha$ we get
 \begin{align}\label{eq:Kaluza-kline_asymptotic_integrand}
    B^{a(D)}_{n,j} 
    &\sim \frac{c_{j,\delta_0}}{n^{\delta_0+1}}(-1)^{j+1}{\oint_{u=0}} \frac{du}{2 \pi i}\frac{ e^{u(3a-\delta+n)}}{ (e^u-1)^{n+2a}}u^{-\delta_0-1}i^{\delta_0-1}
    \begin{cases}
        \cos(u (n{+}\delta))~~~\quad \text{for }\delta_0\text{ odd}\\
       - i \sin (u(n{+}\delta))\quad \text{for }\delta_0\text{ even}
    \end{cases}
\end{align}
for $c_{j,\delta_0}{=}2^{3\delta_0+j+3}(\delta_0+j)!$. Inspired by the methods outlined in \cite{flajolet2009analytic}, which involves using the saddle point method and Hankel deforming the above-closed contour to localize the integral, we evaluate the coefficients to
\begin{equation}
    B^{a(D)}_{n,j}  \sim (-1)^{j+n} \frac{2^{\frac{n}{2}}e^{\frac{\pi}{4}n}}{n^{\frac{D-1}{2}-a}}
       \begin{cases}
        (-1)^{\frac{D+2}{4}}\cos\theta_a(n,\delta,D)\quad \text{for }D \in \{6,10,\dots\}\\
       (-1)^{\frac{D}{4}} \sin\theta_a(n,\delta,D)\qquad \text{for }D \in \{4,8,\dots\}
    \end{cases}
    \label{eq:KK_asymptotic_formula}
\end{equation}
for $\theta_a(n,\delta,D) \equiv \frac{1}{8} \left(\pi  (2 \delta-6n+5)+4 (D-2) \tan ^{-1}\left(\frac{\pi }{\log 4}\right)+4 (n+\delta) \log 2-14a\pi\right)$ up to a positive proportionality constant independent of $n$. We have put the details of the derivation of this formula in Appendix \ref{app:KK-asymptotic_formula_derivation}, and we have tested this formula numerically, finding good agreement for $j\lesssim  \frac{n}{10}$. We see an infinite family of negative coefficients for odd spin resonances at level $n$. This demonstrates the non-unitarity of this model.

\subsection{Asymptotically non-linear Regge spectrum }\label{sec:unitarity_non_linear_regge}

\begin{figure}[ht!]
    \centering
\includegraphics[scale=0.6]{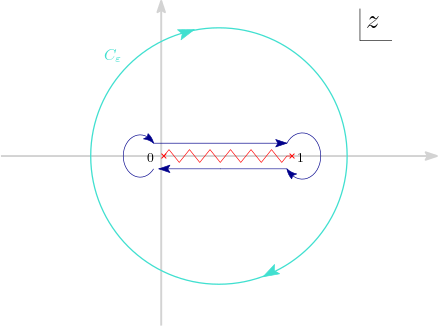}
    \caption{\small Integration contour essential to derive a particular contour representation for the residue of a general bespoke amplitude. The contribution along the small \color{blue}blue \color{black} circles can be shown to vanish for $\text{Re}(\nu(t)),\text{Re}(\nu(s))< -1-a$; the same can be proven for the opposite inequalities using Euler's reflection identity. Going around the branch cut $[0,1]$ denoted by the red jagged line one picks up the phase $e^{-2\pi i \nu_{\alpha}(s)}$. Due to Cauchy's theorem, this is equivalent to the \color{cyan} cyan \color{black} closed contour. After taking the residue, the analyticity of the integrand of \eqref{eq:residue_contour_repn} implies that the branch cut disappears, and the branch point at the origin becomes a pole of order $n+2a$.}
    \label{fig:enter-label}
\end{figure}

We now turn our attention towards studying the unitarity of general bespoke amplitudes with asymptotically non-linear Regge behavior, i.e.~those with $|P|>|Q|+ 1$. We write the spectral curve as
\begin{equation}
    f(\mu,\nu)=P(v)-\mu Q(\nu) = \sum_{m=0}^{|P|}p_m\nu^m-\mu\sum_{m=0}^{|Q|}q_m\nu^m\,.
\end{equation}
We begin by deriving a general expression for the residue as a contour integral for the  bespoke amplitude in its worldsheet integral form
\begin{align}
    A^{a}(s,t)=-\sum_{\alpha,\beta}(\nu_{\alpha}(s))^{a-1}\int_0^1  z^{-\nu_{\alpha}(s)-2a}(1-z)^{-\nu_{\beta}(t)-1-a} dz\,.
\end{align}
By considering a Cauchy's contour deformation argument (see figure \ref{fig:enter-label}), one can see that for an integrand that has the form of $A^{a}(s,t)$, one can take
\begin{equation}
    \int_{0}^{1}dz = \oint_{C_{\varepsilon}}\frac{dz}{e^{-2\pi i \nu_{\alpha}(s)}-1}
\end{equation}
where $C_{\varepsilon}$ is a closed contour of size $\varepsilon>1$. Using the above, we obtain the residue at the $n$-th resonance as a single sum over roots:
\begin{equation}
    R^{a}_{\text{bespoke},n}(t)=\sum_{\beta}\frac{n^{a-1}}{(\nu^{-1}_{\alpha^*}(n))'}\oint_{z=0}\frac{dz}{2\pi i z^{n+2a}}(1-z)^{-\nu_{\beta}(t)-1-a}~\label{eq:residue_contour_repn}\,,
\end{equation}
where owing to the analyticity of the integrand of the above integral, we can actually take $C_{\varepsilon}$ to wind around the origin for any $\varepsilon>0$. 
We have also localized the $\alpha$ sum at some $\alpha^*$ such that $\nu_{\alpha^*}(s_*)=n$\footnote{This, of course, hinges on the assumption that the roots are not degenerate, which is a necessary criterion for the Galois sum to be well defined.}.  We begin our exploration with the monomial spectral curve of degree $m$, given as \begin{equation}
    f(\mu,\nu)=(\nu+\delta)^m-\mu\,,
\end{equation}
solutions of which are
\begin{equation*}
    \nu_{p}(t)=-\delta+t^{\frac{1}{m}}e^{\frac{2\pi ip}{m}}\,,\quad p=1,\,2,\, ...,\,m\,.\end{equation*}
After further substitutions $t=((n+\delta)^m+4a)\frac{x-1}{2}$ and $z=1-e^{-u}$ in \eqref{eq:residue_contour_repn} we have

\begin{align}
  R^{a}_{n}(y)
     &=J_{m,n,\delta}\oint_{u=0}\frac{du}{2\pi i } \frac{e^{u(-\delta+n+3a)}}{(e^u{-}1)^{n+2a}}\sum_{p=1}^{m}e^{u((n+\delta)^{m}+4a)^{\frac{1}{m}}ye^{\frac{i \pi (2p+1)}{m}}}
\end{align}
 for $y=\left(\frac{x-1}{2}\right)^{\frac{1}{m}}e^{\frac{-i\pi}{m}}$. The additional phase in the definition of $y$ is added to make it a real number in the physical regime. We also define $J_{m,n,\delta}=m(n+\delta)^{m-1}n^{a-1}$ and this factor will be dropped for simplicity. 
Therefore, the partial wave coefficient of the amplitude will be
\begin{align}\label{differential_operator_on_exponential}
    B_{n,j}^{a(D)}&=\frac{2^{2\delta_0+j+1}}{m^{j-1}}(-1)^{j}\oint_{u=0}\frac{du}{2\pi i }\frac{e^{u(-\delta+n+3a)}}{(e^u{-}1)^{n+2a}}\mathcal{H}^{a,\delta_0}_{n,j,m}(u,\delta)\,,\nonumber \\
\text{where}~~\mathcal{H}^{a,\delta_0}_{n,j,m}(u,\delta) &= \sum_{p=1}^{m}\int_0^1dyy^{m-1}\left[y^m(1-y^m)\right]^{j+\delta_0}(y^{1-m}\partial_y)^j e^{A^{a,p}_{n,m}(u,\delta)y}\,,\nonumber\\
\text{and}~~A^{a,p}_{n,m}(u,\delta) &\equiv u((n+\delta)^m+4a)^{\frac{1}{m}}e^{\frac{i \pi (2p+1)}{m}}\,.
\end{align}
Later on, for the sake of simplicity of the notation, we will define $A^{a,p}_{n,m}(u,\delta) \equiv A$.

We are now set to prove the claim that for any asymptotically non-linear  spectral curve $f(\mu,\nu) = P(\nu)-\mu Q(\nu)$ the bespoke amplitude
 $A_{\text{bespoke}}$ is non-unitary. Our approach will be to demonstrate that there exists an infinite family of negative coefficients $B^{a(D)}_{n,j}$ in the large $n$ and fixed $j$ limit. It is enough to show that the leading behavior in the large $n$ limit is controlled by the highest power monomial in the spectral curve. To see this, consider a spectral curve whose
 locus of zeroes is described by a degree $|P|-|Q|$ polynomial of the form
\begin{equation}
    \mu(\nu) = P(\nu)/Q(\nu) = (\nu+\delta)^{|P|-|Q|}+\epsilon(\nu+\delta)^{|P|-|Q|-1}\qquad\text{for}~~\epsilon \in \mathbb{R}\,. \label{eq:spectral_curve_monomial}
\end{equation}
Locally near the resonances $\nu_{\alpha}(s)=n$, the roots of the spectral curve as a function of the Mandelstam variable $t$ are then given as the following infinite series in the parameter $\epsilon$:
\begin{equation}
    \nu_{\alpha}(x) = \sum_{k=0}^{\infty}a_{n,\alpha}^{(k)}(x,\delta)\epsilon^k\,, \qquad~~\text{for}~~\alpha\in \mathbb{Z}_{|P|-|Q|}
\end{equation}
where $a_{n,\alpha}^{(0)} = -\delta+\left(\frac{x-1}{2}\right)^{\frac{1}{|P|-|Q|}}e^{\frac{2\pi i \alpha}{|P|-|Q|}}(n+\delta)$ and $x = \cos{\theta}$ as before. Moreover, 
we see that the coefficients of the above series for $k\neq 0$ have the behavior $a_{n,\alpha}^{(k)}\propto \frac{1}{(n+\delta)^{k-1}}$ in the large $n$ limit. As a result in the limit $n\to \infty$ the $a_{n,\alpha}^{(0)}$ term dominates the above series, which arises from the degree $|P|-|Q|$ monomial in the spectral curve \eqref{eq:spectral_curve_monomial}. The parameter $\epsilon$ can be taken to be much smaller than $n$, so this argument holds generally for any arbitrary polynomial. Hence, the analysis for the monomial spectrum $f(\mu,\nu) = (\nu+\delta)^{|P|-|Q|}-\mu$, studied in this section, is applicable to any degree $|P|-|Q|$ polynomial for sufficiently large $n$. 

In order to derive an asymptotic form from (\ref{differential_operator_on_exponential}), we make the change of variable $w=y^m$. Then, applying the partial derivatives via Cauchy's integral formula and taking the $|A| \to \infty$ limit, we get
\begin{align}
    (m\partial_w)^j e^{Aw^{\frac{1}{m}}}&=(-1)^{j}m^j j!\oint_{t=0} \frac{dt}{2\pi i ~t^{j+1}}e^{A(w-t)^{\frac{1}{m}}}\nonumber\\
    &\sim (-1)^{j}m^jA^{jm}j!\oint_{t=0}\frac{dt'}{2\pi i ~t'^{j+1}}e^{Ay-\frac{t'}{mA^{m-1}y^{m-1}}}\\
    &=\frac{A^{j}e^{Ay}}{y^{j(m-1)}}~~~\text{for }~m>1\,,\nonumber
\end{align}
after which the integral over $y$ can be evaluated to
\begin{equation}
    \int_0^1dy^m~[y^m(1-y^m)]^{j+\delta_0}(y^{1-m}\partial_y)^je^{Ay}\sim\frac{m^{j+\delta_0+1}(j+\delta_0)!e^A}{A^{\delta_0+1}}~.
\end{equation}
For $m\neq 1$ the asymptotic form of the partial wave coefficient is
\begin{align}
    B_{n,j}^{a(D)}
    \sim \frac{2^{D-3+j}m^{\frac{D}{2}}\Gamma\left(j+\frac{D-2}{2}\right)}{n^{\frac{D+2}{2}-a-m}}(-1)^{j}\beta^{a(D)}_{n,m}(\delta)\,
    \end{align}
    where
    \begin{align}
   \beta^{a(D)}_{n,m}(\delta) = \sum_{p=1}^m e^{-\frac{i \pi (2p+1)(D-2)}{2m}}\oint_{u=0}\frac{du}{2\pi i}\frac{e^{u(-\delta+n+3a+(n+\delta)e^{\frac{i \pi (2p+1)}{m}})}}{(e^u{-}1)^{n+2a}u^{\frac{D-2}{2}}}\,.
\end{align}
Note that $\beta^{a(D)}_{n,m}(\delta)$ is completely independent of the spin $j$, but there is a sign flipping prefactor $(-1)^j$. Furthermore, our numerical tests have revealed that $\beta^{a(D)}_{n,m}(\delta)$ is non-zero for generic values of the corresponding parameters for large enough $n$. Thus, we see that at large $n$, we have an infinite number of negative partial wave coefficients, establishing the non-unitarity of these models. 

We emphasize that non-unitary holds for all bespoke models with asymptotically non-linear Regge spectra, i.e.~$|P|>|Q|+1$. This result is consistent with the no-go theorem of \cite{Caron-Huot:2016icg}, where the authors argued that any theory with at least one higher spin particle in its spectrum must have an asymptotically linear Regge trajectory\footnote{Theories with accumulation points are exceptions; some notable examples of which are the Coon amplitude \cite{Coon:1969yw,Baker:1970vxk} and some of the variants recently studied by Cheung and Remmen \cite{Cheung:2022mkw,Cheung:2023adk,Geiser:2022exp}. However, investigations have revealed non-unitarity of such amplitudes in their accumulation point spectra \cite{Jepsen:2023sia}.}. This was argued by analyzing the distribution of excess zeros in the asymptotic regime subjected to the constraint of unitarity, whereas we have reached this conclusion by showing that the exchange of sufficiently heavy resonances at fixed spin violates positivity.

\subsection{Asymptotically linear Regge spectrum}\label{sec:unitarity_linear_regge}

We now turn our attention towards studying the unitarity of general bespoke amplitudes with asymptotically linear Regge behavior, i.e.~those with $|P|=|Q|+ 1$. 

\subsubsection{String mass gap and unitarity \label{sec:string_mass_gap}}

We begin by considering the simplest example which is the Veneziano amplitude with a mass gap $\delta$. The spectral curve is then
\begin{align}
f(\mu,\nu) = \nu+\delta-\mu = 0
\end{align}
and it has a single root $\nu_{1} = \mu-\delta$.
Then, the corresponding generalized bespoke amplitude in its integral representation is given as
\begin{equation}
    A^{a}_{\text{mass-gap}}(s,t) = -(s-\delta)^{a-1}\int_0^1dz~z^{-s+\delta-2a}(1-z)^{-t+\delta-1-a}\,.\label{eq:a_deformed_Veneziano}
\end{equation}
The residue of the $n$-th resonance is 
\begin{align}
    R^a_n
    &=\frac{1}{n^{1-a}}\oint_{u=0}\frac{du}{2\pi i}\frac{e^{u(-\frac{3}{2}\delta+\frac{1}{2}n+a)}e^{u(n+\delta+4a)\frac{x}{2}}}{(e^u-1)^{n+2a}}\,.
\end{align}
As in the previous sections we define $x=\cos\theta$ and $t=s\left(\frac{x-1}{2}\right)=(n+\delta)\frac{x-1}{2}$. Following (\ref{partial_wave_coefficient_projection}), we project the residues on the Gegenbauer polynomials to obtain the partial wave coefficients
\begin{align}
    B_{n,j}^{a(D)}
    &=\frac{1}{n^{1-a}}\left(\frac{n+\delta+4a}{2}\right)^j\oint_{u=0}\frac{du}{2\pi i}\frac{e^{u(-\frac{3}{2}\delta+\frac{1}{2}n+a)}u^j}{(e^u-1)^{n+2a}}H_{j+\delta_0}\left(u\frac{n+\delta+4a}{2}\right)\left[1+(-1)^{1+n+j}e^{3u\delta}\right],
\end{align}
where we have dropped the positive constants $\alpha_{j,D}$ and $n_{j,D}$ coming from the Rodrigues formula and we defined the function $H_J(\alpha)=(1-\partial_{\alpha}^2)^J\left(\frac{e^{\alpha}}{\alpha}\right)$. 
Utilizing \eqref{eq:diagonalisation_identity} and applying Cauchy's integral formula, we can now derive a double contour integral representation for the partial wave coefficients:
\begin{align}
   B^{a(D)}_{n,j}(\delta) &\propto \oint_{u=0}\frac{du}{2\pi i}\oint_{v=0}\frac{dv}{2\pi i}\frac{e^{u(n-\delta+3a)}e^{v(2\delta+n+a)}(u-v)^j}{(uv)^{j+\delta_0+1}(e^{u}-e^v)^{n+2a}}\left[1+(-1)^{1+n+j}e^{3(u-v)\delta}\right].
   \end{align}
This is the mass-gap analog of the double contour expression given in \cite{Arkani-Hamed:2022gsa}. Next, consider substituting $u=\log{(1-x)}$ and $v=\log{(1-y)}$ to put the above into the form
\begin{align}
 B_{n,j}^{a(D)}(\delta)&\propto\oint_{x=0}\frac{dx}{2\pi i}\oint_{y=0}\frac{dy}{2\pi i}\frac{1+(-1)^{1+n+j}(1-x)^{-3\delta}(1-y)^{3\delta}}{(1-x)^{1-\delta+a}(1-y)^{1+2\delta+a}(x-y)^{n-j+2a}}\nonumber\\
    &\qquad\qquad\qquad\qquad\times\frac{1}{(\log(1-x)\log(1-y))^{\delta_0+1}}\left(\frac{\frac{1}{\log(1-x)}-\frac{1}{\log(1-y)}}{x-y}\right)^j.
\end{align}
 Following the footsteps of \cite{Arkani-Hamed:2022gsa}, we carry out the $y$ contour integration in the large $n$ fixed $j$ limit, and make the change of variables $x=1+\frac{\tau}{n}$ to obtain the following asymptotic form of the integral:
\begin{align}
   B^{a(D)}_{n,j}(\delta) &\propto \frac{n^{j+\delta_0-\delta+a}}{(\log n)^{\delta_0+1}}\int_{\mathcal{H}}\frac{d\tau}{2\pi i}e^{-\tau}(-\tau)^{-(1-\delta+a)}\left[1+(-1)^{1+n+j}\left(-\frac{\tau}{n}\right)^{-3\delta}\right]
\end{align}
where we considered a deformation of the $x$ closed contour to the Hankel contour $\mathcal{H}$ (see figure \ref{fig:Circle_to_Hankel} in Appendix \ref{app:KK-asymptotic_formula_derivation}). Upon further integration, we finally get the full asymptotic expression for the partial wave coefficients:
\begin{align}
    \label{a-deformed_asymptotic_linear_massgap_bound}
    &B_{n,j}^{a(D)}(\delta)\sim \frac{\mathcal{C}^{D}_{j}}{n^{\delta+\frac{D}{2}-2a}(\log n)^{\frac{D-2}{2}}}\left[\frac{1}{\Gamma(1-\delta+a)}-(-1)^{n+j}\frac{n^{3\delta}}{\Gamma(1+2\delta+a)}\right],\nonumber\\
    &\text{where}~~\mathcal{C}^{D}_{j}\equiv 2^{j+D-3}\Gamma\left(j+\frac{D-2}{2}\right)\alpha_{j,D}n_{j,D}\,.
\end{align}
 which we arrive at utilizing the Hankel contour analytic continuation of the gamma function. In the last expression, for $\delta>0$ the second term will dominate at large $n$, so that the partial wave coefficients will alternate signs as we vary spin at a given resonance level. 
 
 Therefore, we conclude that unitarity requires the mass gap to satisfy the bound $\delta \leq 0$. 
 
\subsubsection{Post-linear Regge spectrum}\label{sec:post_linear_regge_unitarity}

Finally, we consider a bespoke model with a general asymptotically linear Regge spectrum based on a spectral curve with $|P| = |Q| + 1 > 1$. The locus of solutions to the spectral curve was parameterized in  \cite{Cheung:2023uwn} as
\begin{equation}
    \mu(\nu) = \nu+\delta+\kappa_1+\frac{\kappa_2}{\nu+\delta}+\frac{\kappa_3}{(\nu+\delta)^2}+\dots+\frac{\kappa_{|P|}}{(\nu+\delta)^{|P|-1}}\,,\label{eq:post_linear_regge}
\end{equation}
which is a finite series expansion in inverse powers of $\nu+\delta$. We will refer to the $\kappa_i$  for $i \geq 1$ as post-Regge coefficients. The above spectral curve allows for customizable asymptotic corrections to the linear Regge behavior of string theory. We now study the constraints placed on these coefficients by demanding the positivity of partial wave coefficients in the large-$n$ limit for fixed $j$. To achieve this, we first prove an important claim regarding the post-Regge expansion of the bespoke amplitude in this limit.

Consider the spectral curve defined by a degree $|P|-|Q|$ polynomial of the form
\begin{equation}
    \mu(\nu) = P(\nu)/Q(\nu) = (\nu+\delta)^{|P|-|Q|}+\epsilon_1(\nu+\delta)^{|P|-|Q|-1}+\epsilon_2(\nu+\delta)^{|P|-|Q|-2}~~~\text{for}~~\epsilon_{1,2}\in \mathbb{R}\,, \label{eq:spectral_curve_post_regge}
\end{equation}
where $\epsilon_{1}$ and $\epsilon_{2}$ are coefficients, which are the analogs of the leading and sub-leading contributions to the post-Regge spectral curve \eqref{eq:post_linear_regge} respectively. For such a spectral curve, around the poles $\nu_{\alpha}(s)=n$, the roots $\nu_{\alpha}(t)$ will be given as an infinite series in the post-Regge coefficients
\begin{equation}
    \nu_{\alpha}(x) = \sum_{k,\ell\geq 0}a_{n,\alpha}^{(k,\ell)}(x,\delta)\epsilon_1^k\epsilon_2^{\ell}~, ~~~\text{for}~~\alpha\in \mathbb{Z}_{|P|-|Q|}
\end{equation}
where $a_{n,\alpha}^{(0,0)} = -\delta+\left(\frac{x-1}{2}\right)^{\frac{1}{|P|-|Q|}}e^{\frac{2\pi i \alpha}{|P|-|Q|}}(n+\delta)$ and $x = \cos{\theta}$ as before. Moreover, in the relevant limit,
we have $
    a_{n,\alpha}^{(k,0)} \propto \frac{1}{(n+\delta)^{k-1}}~,a_{n,\alpha}^{(0,\ell)}\propto \frac{1}{(n+\delta)^{2\ell-1}}~,~\text{and}~a_{\alpha}^{(k\neq 0,\ell \neq 0)}\propto \frac{1}{(n+\delta)^{\ell+k}}~.
$
 As a result, the $a_\alpha^{(0,0)}$ term dominates the above series, which is simply the root of the degree $|P|-|Q|$ monomial in the spectral curve \eqref{eq:spectral_curve_post_regge}. Moreover, at any given order in the Taylor series of $\epsilon_1$ and $\epsilon_2$,  the sub-leading order coefficients in the post-Regge expansion i.e. $a^{(0,\ell \neq 0)}_{\alpha}$ will be dominated by the leading coefficients, i.e. $a^{(k,0)}_{\alpha}$. Therefore, in the limit $n\to \infty$ there is a hierarchy for these terms based on their large $n$ behaviors. Again, as in section \ref{sec:unitarity_non_linear_regge}, we expect this argument to hold for an arbitrary spectral curve as long as we are in the regime $n\gg\epsilon_1,\epsilon_2$. This argument suggests that asymptotically in $n$, there should be no essential difference between an asymptotically linear Regge spectrum and a strictly linear Regge spectrum. We provide numerical evidence for this in Appendix \ref{app:numerical_checks-hierarchy}.

This fact enables us to impose non-trivial constraints on the post-linear Regge coefficients in \eqref{eq:post_linear_regge}. We achieve this by utilizing the key result of the previous section regarding the large $n$ limit of the partial wave coefficients of the gapped string theory. In particular, we claim that if $\kappa_1\geq 1-2a$, the corresponding bespoke amplitude is either non-unitary or has diverging high energy behavior. To see this, notice that in the limit $\nu \to \infty$ we have $\mu(\nu) = \nu+\delta+\kappa_1+\mathcal{O}(\frac{\kappa_2}{\nu})$; therefore in the large $n$ limit we expect that the partial wave coefficients arising in this model approach the ones derived from string theory with a mass gap $\delta+\kappa_1$. 
 From the analysis in the previous section we have that this mass gap must satisfy $\kappa_1+\delta\leq 0$. In addition, from the constraint that guarantees convergence of high energy asymptotics, namely \eqref{eq:hard_scattering_constraint}, we must also have $\delta > 2a-1$; together these imply that $\kappa_1<1-2a$. 
 
 In the rest of the section, we demonstrate the hierarchy of post-Regge terms and the constraint on $\kappa_1$ analytically. We achieve this by explicitly computing the sub-leading corrections to the large $n$ behavior of the partial wave coefficients for such models.
 For the sake of our analysis, it is sufficient to demonstrate this for the monic quadratic spectrum given by
\begin{equation}
    f(\mu,\nu)=(\nu+\delta)^2+\kappa_1(\nu+\delta)+\kappa_2-\mu(\nu+\delta)\,.
\end{equation}
The $n$-th excitation sits at $s_n=n+\delta+\kappa_1+\frac{\kappa_2}{n+\delta}$. The solutions to the spectral function are
\begin{align}
    \nu_{\pm}(\mu)&=\frac{1}{2}\left(-2\delta-\kappa_1+\mu\pm\sqrt{\kappa_1^2-4\kappa_2-2\kappa_1\mu+\mu^2}\right)\nonumber\\
    &\equiv \nu_0(\mu)\pm\Delta(\mu)\,.
\end{align}
Since we are interested in ruling out inconsistent bespoke models, we conduct the analysis for the spacetime dimension $D=4$; any theory that is non-positive in $D=4$ is necessarily non-positive in any higher dimension. Further, we restrict ourselves to the superstring case i.e. $a=0$, since the derivation for the bosonic case should follow identically. The residue integral then takes the form
\begin{equation}
    R_n(t) = 2\nu_{+}'(s_n)\oint_{u=0}\frac{du}{2\pi i}\frac{e^{u(n+\nu_0(t))}}{(e^u-1)^n}\cosh{u\Delta(t)}
\end{equation}
where as before we have $t= \frac{s_n}{2}(x-1)$.
Then the partial wave coefficients are\footnote{Here, as we aim to conduct an asymptotic analysis in $n$, we have rescaled the partial wave coefficients so as to eliminate all the $n$-independent constants to define $\beta_{n,j}$.} 
\begin{equation}\label{eq:beta_integral_expression}
    \beta_{n,j} 
    = \oint_{u=0}\frac{du}{2\pi i}\frac{e^{u(n-\delta-\frac{\kappa_1}{2})}}{(e^u-1)^n}I_{n,j}(u)\,,
\end{equation}
for the $u$-dependent integral given by
\begin{equation}
    I_{n,j}(u) \equiv \int_{-1}^{1} dx(1-x^2)^{j}\partial_{x}^j\left[e^{\frac{u s_n}{4}(x-1)}\cosh\left(\frac{u}{2}\sqrt{\left(s_n\frac{x-1}{2}-\kappa_1\right)^2-4\kappa_2}\right)\right].\label{eq:I_integral_u}
\end{equation}
The large-$n$\footnote{By large $n$ we mean in particular $n\gg\kappa_2$, in fact, for the numerical analysis conducted in Appendix \ref{app:numerical_checks-asympototic} we find that for the asymptotic formula to show good convergence with the actual result one needs $\frac{n}{\kappa_2}\gtrsim 10^3$. Moreover, for larger values of $\kappa_1$ and $\delta$ one can also improve on this bound.} and fixed-$j$ asymptotic form for $\beta_{n,j}$ is then
\begin{equation}
    \beta_{n,j}\sim \mathcal{B}_{n,j}^{(0)}(\kappa_1,\delta) +\kappa_2\mathcal{B}_{n,j}^{(1)}(\kappa_1,\delta)\label{eq: asymptotic_formula_kappa_2_correction}
\end{equation}
where
\begin{align}
   &\mathcal{B}_{n,j}^{(0)}(\kappa_1,\delta) \sim 2^j j!\bigg[\delta_{0,j}\frac{n^{-\delta}}{\Gamma(1-\delta)}-(-1)^{j+n}\frac{n^{-1+2\delta+2\kappa_1} (\log n)^{-1}}{\Gamma(1+2\delta+2\kappa_1)}+\frac{n^{-1-\delta-\kappa_1}(\log n)^{-1}}{\Gamma(1-\delta-\kappa_1)}\bigg]\label{eq: asymptotic_formula_kappa_2_correction1}\,,\\
  &\mathcal{B}_{n,j}^{(1)}(\kappa_1,\delta) \sim -2^j j!\frac{n^{-1-\delta}(\log n)^2}{\Gamma(1-\delta)}\label{eq: asymptotic_formula_kappa_2_correction2}\,.
\end{align}
The details of the derivation of this formula are given in Appendix \ref{app: post_regge_integral_detail} and we also present numerical checks in Appendix \ref{app:numerical_checks-asympototic}. From (\ref{eq: asymptotic_formula_kappa_2_correction1}) and (\ref{eq: asymptotic_formula_kappa_2_correction2}), we see that in the region where the amplitude has good UV behavior in the hard scattering limit (i.e. $\delta>-1$), and $\kappa_1>\frac{3}{2}$  the partial wave coefficients are dominated by the second term in \eqref{eq: asymptotic_formula_kappa_2_correction1}, which exhibits the sign flipping behavior. If $\kappa_1<0$, then the last term in \eqref{eq: asymptotic_formula_kappa_2_correction1} always dominates over the sub-leading term (\ref{eq: asymptotic_formula_kappa_2_correction2}), which is also positive. Therefore, in the region $\kappa_1\in [\frac{3}{2},\infty)\cup (-\infty,0]$, the above asymptotic formula provides evidence to support the claim made earlier in this section that the partial wave coefficients at large $n$ are dominated by the leading terms in the post-Regge expansion \eqref{eq:post_linear_regge}. \emph{Moreover, \eqref{eq: asymptotic_formula_kappa_2_correction} establishes non-unitarity in the region $\kappa_1>\frac{3}{2}$ and also hints at unitarity whenever $\kappa_1<0$}.

\begin{figure}[h!]
    \centering\includegraphics[scale=0.5]{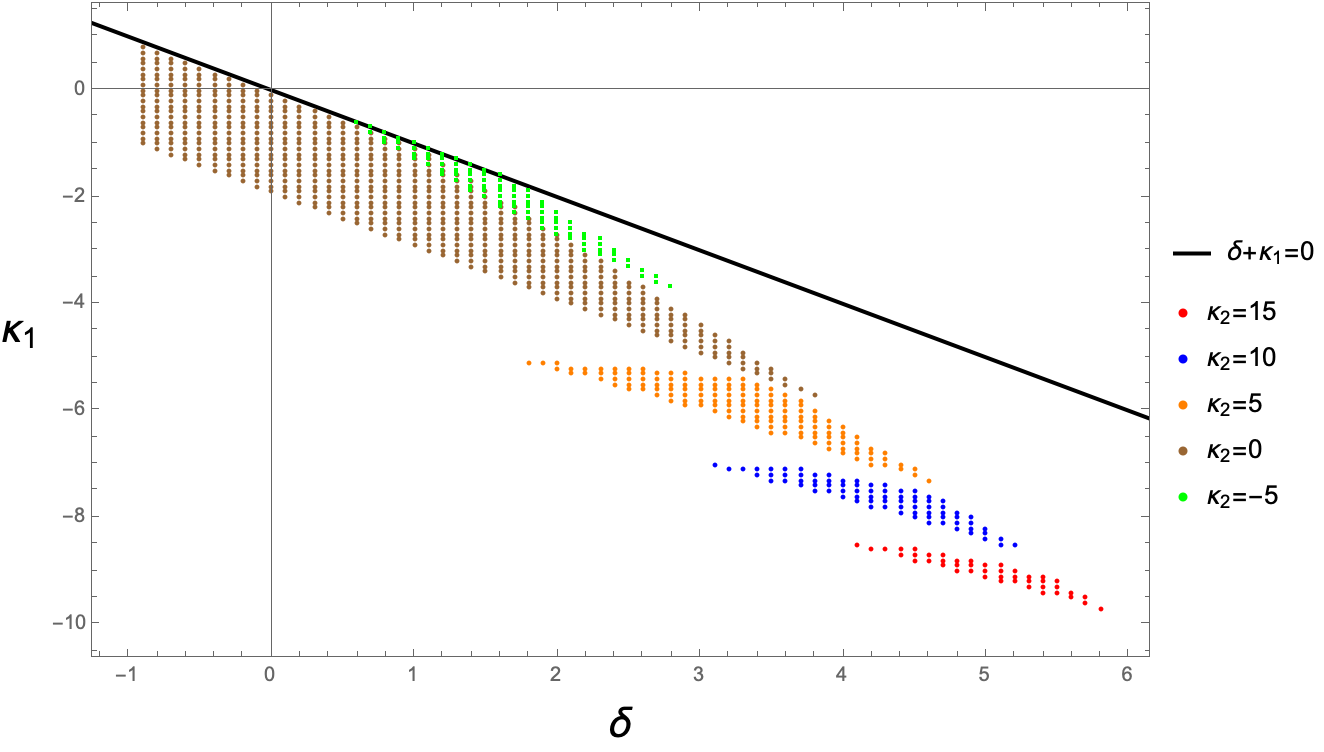}
    \caption{Unitary region in $(\delta,\kappa_1)$ space (checked at $1\leq n\leq 20$), and $\kappa_2$ fixed at various values. There is an overlapping region between $\kappa_2=0$ and $\kappa_2=-5$.}
    \label{fig:small_n_analysis}
\end{figure}

    \emph{Our analysis is inconclusive about non-unitarity in the region $\kappa_1\in (1,\frac{3}{2})$}, where the sub-leading term \eqref{eq: asymptotic_formula_kappa_2_correction2} dominates over the leading terms \eqref{eq: asymptotic_formula_kappa_2_correction1},  and thus the asymptotic formula ceases to hold. However, judging by the form of the original integral \eqref{eq:I_integral_u}, which is analytic in $\kappa_1$, we expect that there exists an appropriate prescription for analytic continuation to this region that we are not aware of at the moment. We, therefore, resort to numerics to provide further evidence of non-unitarity in this region. Scanning over lattice points in $(\delta, \kappa_1)\in ([-1,10],[-10,10])$ space, we plot out the points where we find the partial wave coefficients to be all positive ($1\leq n\leq 20$). The results are shown in figure \ref{fig:small_n_analysis}. 

For all the examples we considered, it was consistently observed that $\kappa_1>1$ leads to non-unitary models, exactly the conclusion earlier derived from the large $n$ analysis. However, not all $\kappa_1<1$ models lead to unitary theories as in figure \ref{fig:small_n_analysis}, there is clearly a smaller sub-region of unitarity for each $\kappa_2$. As a result, our large $n$ analysis provides a necessary bound on the post-Regge coefficient $\kappa_1$ to preserve unitarity but not a sufficient one. Interestingly, the positivity at large $n$ (computed analytically) and that at small $n$ (computed numerically) are independent statements, yet we observe the bound $\kappa_1 \leq 1$ emerging in both analyses. We also notice that all of the unitary regions that we have computed lie below or on the line given by $\delta+\kappa_1=0$. This line is exactly the mass gap bound that we have obtained in section \ref{sec:string_mass_gap}. This shows that the unitary region of the first correction to the post-Regge spectrum is consistent with the mass gap bound of the linear spectrum discussed in section \ref{sec:string_mass_gap}. Furthermore, we observe that the unitary region in $(\delta,\kappa_1)$ is shrinking as $|\kappa_2|$ increases, and this leads us to speculate that there exists a two-sided bound on $\kappa_2$. Indeed, we found no unitary region for points sampled in the regions $\kappa_2\leq-10$ or $\kappa_2\geq35$. It is worth emphasizing, however, that there is no proof to suggest that the region of unitarity in the parameter space $(\delta,\kappa_1,\kappa_2)$ is connected; therefore, we cannot present a bound on $\kappa_2$. Furthermore,  bounds on $\kappa_2$ cannot be extracted from the current large $n$ asymptotic formula; to achieve this, one would have to compute the sub-sub-leading corrections in $n$. We leave such investigations for the future. Finally, note that even though the formula \eqref{eq: asymptotic_formula_kappa_2_correction} is positive for all $\kappa_1<0$, the small $n$ numerical data clearly shows the existence of infinitely many non-unitary points in this region. This further indicates that while large-$n$ studies of positivity are quite efficient at ruling models \emph{out}, they may be poor at ruling models \emph{in}.

\section{Regge Sum Rules and bespoke amplitudes }\label{sec:RSR_bespoke}

In this section, we look at the \textit{Regge Sum Rules} (RSR) and their compatibility with bespoke amplitudes. Most recently, RSR were studied in \cite{Haring:2023zwu}, where the authors used them as constraints on the Wilson coefficients of certain unitary closed and open string amplitudes with general satellite terms. We start by reviewing the RSR and then impose them on our generalized bespoke amplitudes. 

For any four-point amplitude $A(s,t)$, which in general has simple poles in $s,t$ and/or a branch cut along the real axis in the complex  $s$  plane, the RSR constraint implies the following integral equation is satisfied for sufficiently negative $t$:
\begin{equation}
    \int_{-\infty}^{\infty} ds'~ (s')^{n}~\text{Disc}_sA(s',t) = 0~,\qquad\forall~ n \in \mathbb{N} \label{eq:rsr}.
\end{equation}
This primarily comes from the constraint of \textit{superpolynomial boundedness} or \textit{superpolynomial softness}, i.e.~the amplitude in the large and negative $s$ limit falls off to zero faster than any power law.
\begin{figure}[h!]
    \centering
\includegraphics[scale=0.6]{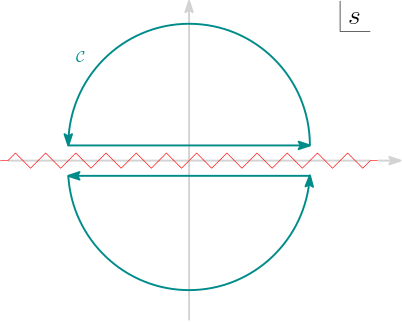}
\caption{A Cauchy-like contour deformation argument in the complex $s$ plane is used to derive the so-called Regge Sum Rules. The branch cut is taken to run along the real line without loss of generality.}
\label{fig:RSR_closed_contour}
\end{figure}

 More concretely, integrating the amplitude along the closed contour $\mathcal{C}$ shown in figure \ref{fig:RSR_closed_contour} will enclose simple poles in either the upper or the lower half of the $s$-plane and thus by virtue of Cauchy's theorem we have 
\begin{equation}
    \oint_{\mathcal{C}} ds'~(s')^n~A(s',t) = 0 \label{eq:cauchy_rsr}
\end{equation}
for any positive integer $n$. But note that $\mathcal{C} = \mathcal{C}_{\infty}+(-\infty+i\epsilon,\infty+i\epsilon)+(\infty-i\epsilon,-\infty-i\epsilon)$, where $\mathcal{C}_{\infty}$ is the large circular contour enclosing the complex plane, the integral along which is given as 
\begin{equation}
    \int_{\mathcal{C}_{\infty}}ds'~(s')^n~A(s',t)\sim \int_{\mathcal{C}_{\infty}}\frac{ds'}{s'}~(s')^{n+j(t)+1}
\end{equation}
where we have made use of the Regge behavior of the amplitude $A(s,t) \sim s^{j(t)}$ as $|s|\to \infty$ with $t$ held fixed. Now if $j(t)<-n-1$ we have that the above integral vanishes:
\begin{equation}
    \int_{\mathcal{C}_{\infty}}ds'~(s')^n~A(s',t)= 0 \qquad \text{if} ~~ j(t)<-n-1\,. \label{eq: RSR_constraint}\
\end{equation}
Taking $\epsilon \to 0$ on the real line contours gives us the integral of the discontinuity:
\begin{align}
   \left( \int_{-\infty+i\epsilon}^{\infty+i\epsilon}+\int_{\infty-i\epsilon}^{-\infty-i\epsilon}\right)ds'~(s')^nA(s',t) &= \int_{-\infty}^{\infty} ds'~(s')^n(A(s'+i\epsilon)-A(s'-i\epsilon))\nonumber\\ &= \int_{-\infty}^{\infty} ds'~(s')^n\text{Disc}_sA(s',t)~,
\end{align}
and thus using the above results combined with the fact \eqref{eq:cauchy_rsr} one arrives at the RSR given by \eqref{eq:rsr}. For string theory $j(t)=t$, and thus \eqref{eq: RSR_constraint} is satisfied for all $t<-n-1$, which is always achievable as $t$ can be made arbitrarily negative. As $A_V(s',t)$ is meromorphic, therefore  $\text{Disc}_sA(s',t)= \sum_{k=0}^{\infty}R_{V,k}(t)\delta(s'-k)$. As a consequence, the RSR constraint simply implies that the sum of all residues vanishes, which is obvious by virtue of Cauchy's theorem for any $n$ in \eqref{eq:rsr}. 

We will now look for bespoke amplitudes other than string theory that satisfy RSR and are also unitary and showcase consistent high-energy behavior. First assume $|P| \neq 1$. Then, in the Regge limit, we know that 
 \begin{equation}
         A^a_{\infty}(t) \sim \sum_{\beta} \left(s^{\nu_{\beta}(t)+2a-1}+\sum_{\alpha \neq 0}(\nu_{\alpha}^{(Q)})^{\nu_{\beta}(t)+2a-1}\right).
     \end{equation}
     Generally it is always possible to guarantee that the roots of the spectral curve can be taken to be arbitrarily negative, i.e.~$j(t) = \text{Re}(\nu_{\beta}(t))<-n-2a$ for any $\beta$ and $n \in \mathbb{N}$, so the first term could be consistent with the constraint \eqref{eq: RSR_constraint}. Then, for the RSR to hold, we must have 
     \begin{equation}
         \sum_{\beta}\sum_{\alpha \neq 0}(\nu_{\alpha}^{(Q)})^{\nu_{\beta}(t)+2a-1+n} = 0~,\qquad \forall ~n\in \mathbb{N}\,.
     \end{equation}
     In general it is not possible for this to be true for all $t$ in the kinematic region defined by the high energy behavior of the amplitude, so we must have that either $Q$ is trivial (i.e.~$|Q|=0$) or $\nu_{\alpha}^{(Q)} = 0$ for all $\alpha = 1,\dots,h-1$, which implies $Q(\sigma) = \sigma^{h-1}$\footnote{We have checked this with numerical trial and error, but we do not claim to have a formal proof of this statement.}. If $|Q|=0$, then the analysis of non-linear Regge spectra in section \ref{sec:unitarity_non_linear_regge} shows that the amplitude violates unitarity. This leaves us with $Q(\sigma) = \sigma^{h-1}$ for $h \neq 1$, which is simply the post-Regge bespoke model that we studied in section \ref{sec:post_linear_regge_unitarity} with vanishing mass gap. 
     
     Tree-level string amplitudes are known to possess exponentially soft behavior and hence satisfy RSR. Furthermore, it was demonstrated by the authors in \cite{Veneziano:2017cks} (see also \cite{Tryon:1972tp}) that pion-pion scattering in large $N$ QCD could only be described by a class of UV finite, dual resonant and unitary amplitudes that are not superpolynomialy soft and do not satisfy RSR. It is, therefore, interesting to chart out the full space of UV finite amplitudes that satisfy RSR. The analysis above enabled us to conclude that a large class of bespoke amplitudes does not satisfy RSR. The same has been shown for the hypergeometric amplitudes \cite{Cheung:2023adk} in \cite{Haring:2023zwu}, which is another class of UV finite, dual resonant and unitary amplitudes. 
     
\section{Discussion}\label{sec:discussion}

 We now reflect on some of the key results of our paper and outline some future directions and open questions. The main results are:
    \begin{itemize}
        \item We generalized the bespoke amplitude to type-I superstring theory. This was enabled by the introduction of parameter $a$, such that for $a=1$ one recovers the bosonic bespoke amplitude, whereas for $a=0$ we obtain the superstring generalization thereof. We explicitly established dual resonance and evaluated the high-energy behavior of this generalization. 
        \item We demonstrated the non-unitarity of bespoke amplitudes with asymptotically non-linear Regge trajectories, which concurs with the no-go results of \cite{Caron-Huot:2016icg}. This was achieved by studying the positivity of partial wave coefficients at large $n$ for fixed spin $j$.
        \item  In the same manner, we showed that unitarity imposes $\delta \leq 0$ on the mass gap $\delta$ for a linear Regge trajectory shifted by a mass gap $\delta$.  In the context of the bespoke Veneziano amplitude ($a=1$), this is in accord with the constraint $\alpha_0 \ge -1$ evident in figure~1 of \cite{Arkani-Hamed:2023jwn}, which was derived by studying the positivity of light resonances (small $n$). 
        \item For bespoke amplitudes with non-linear but asymptotically linear Regge trajectories, we showed that requiring positivity of the partial wave expansion imposes a non-trivial constraint on the leading post-Regge coefficient, namely $\kappa_1< 1-2a$. 
        \item Finally, we demonstrated that even though most bespoke amplitudes seem to violate Regge Sum Rules, there exists an infinite sub-class of unitary bespoke models consistent with these rules; these are the post-Regge bespoke models with a vanishing mass gap and asymptotically linear Regge trajectories.
    \end{itemize}
    
    It is important to stress the fruitfulness of studying the positivity of partial wave coefficients in the large $n$ limit. This approach has facilitated ruling out of a large class of bespoke amplitudes, which earlier could not be negated on the basis of the positivity of some of the low-lying coefficients \cite{Cheung:2023uwn}. The study of large-$n$ asymptotics of $B^D_{n,j}$ was achievable due to its particularly elegant contour-integral representation, which allowed the integrand to be localized around its saddle points in the limit via a suitable contour deformation. Fundamentally, this was possible because the bespoke amplitudes admit a worldsheet representation. However, for amplitudes that do not admit such representations, such as the Coon amplitude, it is not clear how to carry out an analytic large-$n$ analysis, so one is limited to numerical methods.
    
    It is also important to study other, possibly stronger, constraints that one can impose on the $S$-matrix beyond those of the positivity of the four-point partial wave coefficients. More recently, there has been some advancement in the direction of higher-point $S$-matrix bootstrap \cite{Caron-Huot:2023ikn} and higher-point factorization \cite{Geiser:2023qqq,Arkani-Hamed:2023jwn}.  To compare the strength of higher-point factorization against partial wave unitarity, consider the following toy example of a deformed Veneziano amplitude first suggested in \cite{Arkani-Hamed:2023jwn}:
    \begin{equation}
        \mathcal{A}(s,t) = \int_0^1dz~F(z)~z^{-s-1}(1-z)^{-t-1}\,,\label{eq:P_deformed_amp}
    \end{equation}
    where $F$ is an arbitrary regular function which, due to crossing, is forced to satisfy $F(z)= F(1-z)$. By Taylor expanding $F(z)$ and integrating term by term, the above can be shown to be equivalent to a sum over satellite terms 
    \begin{equation}
        \mathcal{A}(s,t) = \sum_{i,j}a_{i,j}\frac{\Gamma(-s+i)\Gamma(-t+j)}{\Gamma(-s-t+i+j)} \label{eq:satellites_F_deformed_amp}
    \end{equation}
    where $a_{i,j}=a_{j,i}$ are the Taylor coefficients of $F(z)$. This was first proposed by Gross in 1969 \cite{Gross:1969db}, in the context of consistent higher-point factorization of string amplitudes. For the special case when $F$ is a polynomial, these amplitudes could be ruled out on the grounds of factorization as demonstrated in \cite{Arkani-Hamed:2023jwn}. However, when $F$ is not a polynomial and the sum in (\ref{eq:satellites_F_deformed_amp}) is infinite, for certain special cases when $F$ is an exponential of the cross ratios, the above suffers from degeneracy at massive resonances and hence can no longer be ruled out on the basis of factorization alone. It would be interesting to study the partial wave unitarity of these amplitudes. 
    
    Moreover, in \cite{Haring:2023zwu}, the authors studied a generalization of amplitudes of this type from the perspective of RSR and unitarity, finding various examples that could satisfy both (cf.~Matsuda and Mandelstam amplitudes). One can readily derive the following double contour formula for the partial wave coefficients of (\ref{eq:P_deformed_amp}):
    \begin{equation}
        B^D_{n,j} \propto \oint_{u=0}\frac{du}{2\pi i}\oint_{v=0} \frac{dv}{2\pi i} F(e^{v-u})\frac{(v-u)^j}{(uv)^{j+\delta_0+1}(e^v-e^u)^{n}}
    \end{equation}
    up to some manifestly positive proportionality constant,
    which is similar to the double contour representation in \cite{Arkani-Hamed:2022gsa}. It would be interesting to determine what is the allowed class of functions $F$ that guarantees manifest positivity for a certain range of spacetime dimensions $D$, which class of functions generate positive partial waves in the large $n$ limit, and to compare those with the results of higher-point factorization in \cite{Arkani-Hamed:2023jwn}.

It would also be interesting to study where the space of Wilson coefficients of bespoke amplitudes lies in the EFThedron \cite{Arkani-Hamed:2020blm,Caron-Huot:2020cmc}. More explicitly, working in the primal approach, one may begin with arbitrary polynomials $P$ and $Q$ of fixed degree $h$ and $h-1$ respectively for $h\geq 2$, and obtain the Wilson coefficients of the corresponding bespoke amplitude in the low energy limit. These Wilson coefficients will be functions of the coefficients of the two polynomials. One can then constrain these Wilson coefficients using partial wave unitarity analysis carried out in our work. Since the scope of this paper was limited to constraining the amplitudes in the high-energy regime, this would serve as a complementary low-energy study. 
  
  The limited utility of four-point partial wave unitarity in constraining the space of physical dual resonant models is quite evident from our analysis. It is, therefore, important to develop additional tools and techniques required to conduct studies of the unitarity of higher-point amplitudes. An obvious roadblock here is the absence of a partial wave basis. However, in the past, a few investigations have been conducted in the direction of Gegenbauer polynomials that are a function of multiparticle scattering angles cf.~\cite{PhysRevD.4.489,PhysRevD.4.2260}. 

\acknowledgments

We are grateful to Shounak De, Yu-tin Huang, Gareth Mansfield and
Grant Remmen for helpful discussions. This work was supported in part by the US Department of Energy under contract {DE}-{SC}0010010 Task F (RB, MS, AV), by Simons Investigator Award \#376208 (AV), and by Bershadsky Distinguished Visiting Fellowships at Harvard (MS, AV). Part of this research was conducted using computational resources and services at the Center for Computation and Visualization, Brown University.

\appendix

\section{Details on the derivation of key formulae }\label{app:KK-asymptotic_formula_derivation}
\subsection{Derivation of \eqref{eq:KK_asymptotic_formula}}

In this appendix, we carry out the derivation of \eqref{eq:KK_asymptotic_formula}. We start by substituting $u = \log{(1-x)}$ in \eqref{eq:Kaluza-kline_asymptotic_integrand}, which puts the integral into the form
\begin{equation}
    B^{(D)}_{n,j} \sim J_{n,\delta}\frac{(-1)^{j+\frac{D+2}{4}+n}+1}{n^{\frac{D-2}{2}}}\oint_{x=0}\frac{dx}{2\pi i}\frac{(1-x)^{n-\delta+3a}}{(\log(1-x))^{\frac{D-2}{2}}}\frac{1}{x^{n+2a}}
    \begin{cases}
        \cos( (n{+}\delta)\log(1-x))\quad \text{for }\delta_0\text{ odd}\\
        -i \sin ((n{+}\delta)\log(1-x))\quad \text{for }\delta_0\text{ even}
    \end{cases}
\end{equation}
where we have dropped all proportionality constants independent of $n$. For the sake of computational simplicity, we will also drop the Jacobian $J_{n,\delta}$ and reintroduce it towards the end of the section.

\begin{figure}[h!]
      \centering
      \includegraphics[scale=0.6]{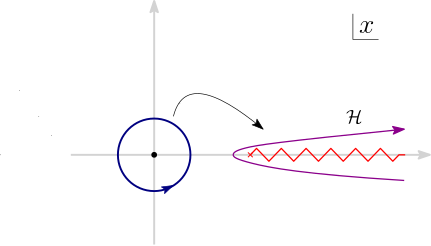}
      \caption{Deformation of the circular contour into Hankel contour.}
      \label{fig:Circle_to_Hankel}
  \end{figure}

Next, following the prescription stated in section VI.2 of \cite{flajolet2009analytic} we deform our circular contour around $x=0$ to the Hankel contour with the branch cut running from $x=1$ to $x=\infty$ as shown in the figure \ref{fig:Circle_to_Hankel}\footnote{Note that in order for this deformation to work we need the contribution from infinity to vanish, which is only true if $\delta>-1$.}.  We now analyze each of the cases individually. 
 
\begin{itemize}
    \item \textbf{The odd $\delta_0$ case}
    \\
       Consider first the further substitution $x=1+\tau$ that shifts the pole to $-1$, and the branch cut to $[0,\infty]$. Then, using Euler's identity, we can bring the integral into the form
        
        \begin{equation}
            B^{(D)}_{n,j} \sim \frac{(-1)^{j+\frac{D-2}{4}-n}}{2\, n^{\frac{D-2}{2}}} \int_{\mathcal{H}}\frac{d\tau}{2\pi i}\frac{\left(-\tau\right)^{s^a_{n}(\delta)}{+}\left(-\tau\right)^{\bar{s}^a_{n}(\delta)}}{(\log(-\tau))^{\frac{D-2}{2}}(1+\tau)^{n+2a}}
        \end{equation}
        where $s^a_{n}(\delta) \equiv n-\delta+3a-i(n{+}\delta)$, and $\bar{s}$ is its complex conjugate. 
        Notice that our integral has the general form 
        \begin{equation*}
            \int_{\mathcal{C}}d\tau ~g(\tau)e^{nf(\tau)}+\text{c.c.}
        \end{equation*}
        where $\mathcal{C}=\mathcal{H}$ and 
        \begin{align*}
            f(\tau) &=  (1+i)\log(-\tau)-\log(1+\tau)\,,~~
            g(\tau) = \frac{(-\tau)^{-\delta(1-i)+3a}}{(\log(-\tau))^{\frac{D-2}{2}}(1+\tau)^{2a}}\,.
        \end{align*}
        By deforming the contour $\mathcal{H}$ along the path of steepest descent $\theta$ passing through the saddle point $\tau_0$ of $f$ and utilizing the saddle point approximation 
        \begin{equation}
            \int_{\mathcal{C}}d\tau g(\tau)e^{nf(\tau)} \sim g(\tau_0)e^{nf(\tau_0)}\sqrt{\frac{2\pi}{n|f''(\tau_0)|}}e^{i\phi}~~~~\text{for}~~\phi =  \frac{(2m-1)\pi-\theta}{2}~,~m\in\mathbb{N} \label{eq:saddle_approximation_formula}
        \end{equation}
        we can localize our integral in the neighborhood of $\tau_0$.
         For us the saddle point is $\tau = i-1 = \sqrt{2}e^{\frac{3\pi i}{4}}$, and furthermore  $\phi =  \frac{7\pi}{8}$ (taking $m=2$ in \eqref{eq:saddle_approximation_formula})\footnote{We choose this as opposed to the negative phase $-\frac{\pi}{8}$ because we work on the principle branch $[0,2\pi)$.}. 
   The deformation of the Hankel contour to the path of steepest descent is visualized in figure \ref{fig:saddle-point-contour-deformation}. This allows us to do the integral and therefore, we finally have 
        \begin{align*}
            B^{(D)}_{n,j} 
            &\sim (-1)^{j+n+\frac{D+2}{4}} \frac{2^{\frac{1}{2}n}e^{\frac{\pi}{4}n}}{n^{\frac{D-1}{2}}}\cos\theta_a(n,\delta,D) 
        \end{align*}
        for the phase
        \begin{align*}    
        \theta_a(n,\delta,D) \equiv \frac{1}{8} \left(\pi  (2 \delta-6n+5)+4 (D-2) \tan ^{-1}\left(\frac{\pi }{\log 4}\right)+4 (n+\delta) \log 2-14a\pi\right).
        \end{align*}
   \begin{figure}
       \centering
       \includegraphics[scale=0.5]{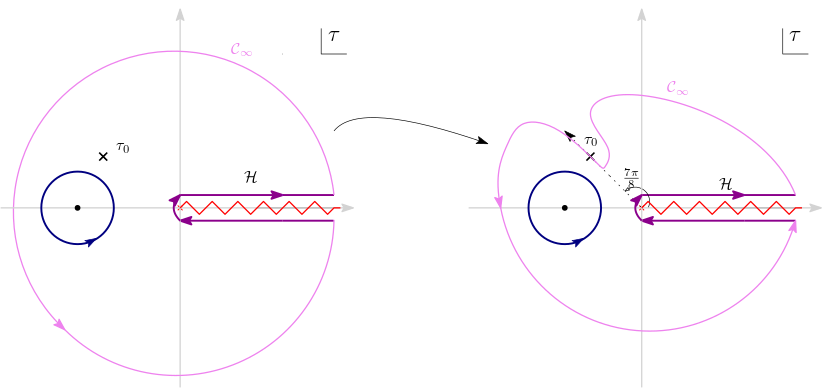}
       \caption{Deformation of the Hankel contour along the path of steepest descent. The contribution from $\mathcal{C}_{\infty}$ vanishes and hence the closed contour is a valid deformation of the Hankel contour.}
       \label{fig:saddle-point-contour-deformation}
   \end{figure}
    
   \item \textbf{The even $\delta_0$ case}
   
   After doing the substitutions mentioned for the odd case, the integral for the even case takes the form
    \begin{equation}
            B^{(D)}_{n,j} \sim \frac{(-1)^{j+\frac{D+2}{4}
            +n}}{2\,n^{\frac{D-2}{2}}}\int_{\mathcal{H}}\frac{d\tau}{2\pi i}\frac{\left(-\tau\right)^{s^a_{n}(\delta)}{-}\left(-\tau\right)^{\bar{s}^a_{n}(\delta)}}{(\log(-\tau))^{\frac{D-2}{2}}(1+\tau)^{n+2a}}\,.
        \end{equation}
    Following the same steps as before leads to the coefficients 
    \begin{align}
          B^{(D)}_{n,j}  &\sim (-1)^{j+n+\frac{D-4}{4}} \frac{2^{\frac{1}{2}n}e^{\frac{\pi}{4}n}}{n^{\frac{D-1}{2}}}\sin\theta_a(n,\delta,D)\,.
        \end{align}
\end{itemize}
 Therefore, the final result reads 
\begin{equation}
    B^{(D)}_{n,j} \sim (-1)^{j+n} \frac{2^{\frac{1}{2}n}e^{\frac{\pi}{4}n}}{n^{\frac{D-1}{2}-a}}
       \begin{cases}
        (-1)^{\frac{D+2}{4}}\cos\left(\theta_a(n,\delta,D)\right)\quad \text{for }D \in \{6,10,\dots\}\\
       (-1)^{\frac{D}{4}} \sin\left(\theta_a(n,\delta,D)\right)\quad \text{for }D \in \{4,8,\dots\}~.
    \end{cases}
\end{equation}
where we have reintroduced the Jacobian as promised earlier in the section and dropped all $n$ independent proportionality positive constants. The above matches the result mentioned in section \ref{sec:KK_unitarity_toy_example}.

\subsection{Derivation of \eqref{eq: asymptotic_formula_kappa_2_correction}}\label{app: post_regge_integral_detail}

Starting from \eqref{eq:I_integral_u} we make the linear substitution $z=-\frac{x-1}{2}$, so that the integral gets transformed into
\begin{align}
    I_{n,j}(u)&=(-1)^{j}2^{j+2\delta_0+1}\int_0^1 dz~z^{j}(1-z)^{j}\partial_z^j\left[e^{-\frac{u s_n z}{2}}\cosh\left(\frac{u}{2}\sqrt{(s_n z+\kappa_1)^2-4\kappa_2}\right)\right].\label{eq:I-integral-1}
\end{align}
Next, we want to consider the limit $s_n\rightarrow\infty$, which means the following approximation for the integrand
\begin{align}
    &\cosh\left(\frac{u}{2}\sqrt{(s_n z+\kappa_1)^2-4\kappa_2}\right)\nonumber\\
    &\qquad\sim\cosh\left(\frac{u (s_n z + \kappa_1)}{2}\right)-\frac{u\kappa_2}{s_n z+\kappa_1}\sinh\left(\frac{u (s_n z + \kappa_1)}{2}\right)+\mathcal{O}\left(\kappa_2/s_n\right)^2~.
\end{align}
Hitting the above with the partial derivatives $\partial_z^j$ then gives us
\begin{align}
    &\partial_z^j\left[e^{\frac{-u s_n z}{2}}\cosh\left(\frac{u}{2}\sqrt{(s_n z+\kappa_1)^2-4\kappa_2}\right)\right]\nonumber\\
    &\sim\frac{1}{2}\left(e^{\frac{u \kappa_1}{2}}\delta_{0,j}+(-1)^{j}(u s_n)^je^{-u s_n z-\frac{u \kappa_1}{2}}-\frac{(-1)^{j}s_n^{j} u \kappa_2}{( s_n z+\kappa_1)^{j+1}}e^{\frac{u\kappa_1}{2}}\gamma(1+j,u (s_n z+\kappa_1))\right)\label{eq:action_of_derivative}
\end{align}
where $\gamma(a;z)$ is the incomplete gamma function, the definition and properties of which have been outlined in Appendix \ref{app:mathematical_addendum_special_functions}. The integral \eqref{eq:I-integral-1} then takes the form
\begin{align}\label{eq:I-integral-2}
    I_{n,j}(u) &\sim 2^{j+2\delta_0}\int_0^1dz\bigg[z^{j}(1-z)^{j}\left(e^{\frac{u \kappa_1}{2}}(-1)^j\delta_{0,j}+(u s_n)^je^{-u s_n z-\frac{u\kappa_1}{2}}\right)\nonumber\\
    &\qquad\qquad\qquad-z^{j}(1-z)^{j}\frac{s_n^j u \kappa_2 e^{\frac{u\kappa_1}{2}}}{(s_n z+\kappa_1)^{j+1}}\gamma(1+j,u (s_n z+\kappa_1))\bigg].
\end{align}
We now integrate the above term by term, with the following results:
\begin{itemize}
    \item $\begin{aligned}
        \int_0^1 dz\ z^j(1-z)^j&=\frac{\Gamma(1+j)^2}{\Gamma(2+2j)}\,,
    \end{aligned}$
    \item 
    $\begin{aligned}
        \int_0^1 dz\ z^j(1-z)^je^{-u s_n z}&=\sqrt{\pi}(-u s_n)^{-\frac{1}{2}-j}e^{\frac{-u s_n}{2}}I_{\frac{1}{2}+j}\left(\frac{-u s_n}{2}\right)\nonumber\\
    &\sim \frac{j!}{(s_nu)^{1+j}}((-1)^{1+j}e^{-u s_n}+1)+\mathcal{O}(s_n^{-2-j})\,,\\
    \end{aligned}$
    \item $\begin{aligned}
         &\int_0^1dz\ z^j(1-z)^j\frac{\gamma(1+j,u(s_n z+\kappa_1))}{(s_n z+\kappa_1)^{j+1}} \sim \frac{j!}{s_n^{j+1}}\left(\frac{(-1)^je^{-u(s_n+\kappa_1)}}{s_n u}+\log s_n\right)+\mathcal{O}(s_n^{-j-2})\,,
    \end{aligned}$
\end{itemize}
where we utilized several identities of the incomplete gamma function in the $s_n \to \infty$ limit while deriving the form of the third integral, and the integral definition and the asymptotic expansion of the Bessel-$I$ function presented in Appendix \ref{app:mathematical_addendum_special_functions} for obtaining the second integral. Plugging these results back into \eqref{eq:I-integral-2} we obtain the approximation of \eqref{eq:I-integral-1}
\begin{align}
    &I_{n,j}(u)\sim 2^{j+2\delta_0}\bigg[\delta_{0,j}e^{\frac{u\kappa_1}{2}}+j!e^{-\frac{u\kappa_1}{2}}\frac{(-1)^{j+1}e^{-u s_n}+1}{s_n u}\nonumber\\
    &\qquad\qquad\qquad\qquad- \kappa_2 ue^{\frac{u\kappa_1}{2}}\frac{j!}{s_n}\left(\frac{(-1)^je^{-u(s_n+\kappa_1)}}{s_n u}+\log s_n\right)
    \bigg]\,.
\end{align}
Next to obtain the partial wave coefficients $\beta_{n,j}$ we perform the relevant $u$ contour integrals in \eqref{eq:beta_integral_expression} which evaluate to the following:
\begin{itemize}
    \item $\begin{aligned}
        &\oint_{u=0}\frac{du}{2\pi i}\frac{e^{u(n-\delta)}}{(e^u-1)^n}\sim\frac{n^{-\delta}}{\Gamma(1-\delta)}\,,
    \end{aligned}$
    \item $\begin{aligned}
        \oint_{u=0}\frac{du}{2\pi i}\frac{e^{u(n-\delta)}u}{(e^u-1)^n }
    \sim \frac{n^{-\delta} \log n}{\Gamma(1-\delta)}\,,
    \end{aligned}$
    \item $\begin{aligned}
        &\oint_{u=0}\frac{du}{2\pi i}\frac{e^{u(n-\delta-\kappa_1-s_n)}}{(e^u-1)^n u}=\frac{1}{n!}B_{n}^{(n)}\left(-2\delta-2\kappa_1-\frac{\kappa_2}{n+\delta}\right)
    \sim\frac{(-1)^n n^{-1+2\delta+2\kappa_1}(n+\kappa_2\log n)}{\Gamma(1+2\delta+2\kappa_1)\log n}\,,
    \end{aligned}$
    \item $\begin{aligned}
        &\oint_{u=0}\frac{du}{2\pi i}\frac{e^{u(n-\delta-\kappa_1)}}{(e^u-1)^n u}=\frac{(-1)^{n}}{n!}B_n^{(n)}(\delta+\kappa_1)\sim\frac{1}{\log n\ n^{\delta+\kappa_1}\Gamma(1-\delta-\kappa_1)}\,,
    \end{aligned}$
    \item $\begin{aligned}
        &\oint_{u=0}\frac{e^{u(n-\delta-s_n-\kappa_1)}}{(e^u-1)^n}=\frac{1}{(n-1)!}B_{n-1}^{(n)}\left(-2\delta-2\kappa_1-\frac{\kappa_2}{n+\delta}\right)\sim \frac{(-1)^{n+1}n^{2\delta+2\kappa_1}}{\Gamma(1+2\delta+2\kappa_1)}\left(1+\frac{\kappa_2\log n}{n }\right).
    \end{aligned}$
    
\end{itemize}
The first two integrals were evaluated using the techniques mentioned in \cite{flajolet2009analytic}. While evaluating the final three integrals, we have utilized the generating function of the Bernoulli polynomials $B^{(n+\nu)}_{n}(z)$ and their large $n$ and fixed $\nu,z$ asymptotic expansion as given in \cite{Temme:1996}, which we also review in Appendix \ref{app:mathematical_addendum_special_functions}. Plugging these results back into \eqref{eq:beta_integral_expression}, we finally obtain the form for the partial wave coefficients presented in section \ref{sec:post_linear_regge_unitarity}:
\begin{align*}
    \beta_{n,j}
    &\sim 2^j j!\bigg[\delta_{0,j}\frac{n^{-\delta}}{\Gamma(1-\delta)}+(-1)^{1+j+n}\frac{n^{-1+2\delta+2\kappa_1} (\log n)^{-1}}{\Gamma(1+2\delta+2\kappa_1)}+\frac{n^{-1-\delta-\kappa_1}(\log n)^{-1}}{\Gamma(1-\delta-\kappa_1)}-\kappa_2 \frac{n^{-1-\delta}(\log n)^2}{\Gamma(1-\delta)}\bigg]~,
\end{align*}
where while deriving the above we dropped several numerical constants.

\section{Numerical sanity checks}

\subsection{Equivalence of asymptotically linear and strictly linear as $n\to \infty$}\label{app:numerical_checks-hierarchy}

Here, we define the \textit{strictly linear} spectral function
\begin{equation}
    f(\mu,\nu)=\nu+\delta+\kappa_1-\mu
\end{equation}
and the \textit{asymptotically linear} spectral function
\begin{equation}
    f(\mu,\nu)=(\nu+\delta)^h+\sum_{i=1}^{h}\kappa_i(\nu+\delta)^{h-i}-\mu(\nu+\delta)^{h-1}\,.
\end{equation}
The root of the strictly linear spectral function is
\begin{equation}
    \nu_{\text{strictly linear}}=\mu-\delta-\kappa_1,
\end{equation}
while the asymptotically linear spectral function has $h$ roots. If we set $\kappa_i=0\ \text{for all } i\geq2$, then the roots of the asymptotically linear spectral function will reduce to one root at $\nu_{\text{strictly linear}}$ with the remaining roots degenerating to $\nu=-\delta$. We call this particular case the \textit{linear} spectral function. 

The residues of the bespoke amplitude (for $a=0$) can be written as
\begin{align}
    R_n(t)=J_n\oint_{u=0}\frac{du}{2\pi i}\frac{e^{u (n+1)}}{(e^u-1)^n}\sum_{\alpha}e^{\nu_\alpha(t)}\,,
\end{align}
where $J_n$ is a Jacobian factor that is independent of spin. Setting $\kappa_i=0\ \text{for all } i\geq2$ will reduce the sum over roots to $\sum_{\alpha}e^{\nu_{\alpha}(t)}=e^{\nu_{\text{strictly linear}}(t)}+(n-1)e^{-\delta}$. This implies that the residue of the amplitude is the strictly linear residue plus a $t$-independent factor. Therefore, asymptotically in $n$, the only difference between the residues of the strictly linear amplitude and the linear amplitude is the spin-0 sector of the partial wave expansion. Therefore, we expect that as we tune $\kappa_2 \to 0$ in the first correction to the post-Regge expansion, we would find that $\beta_{n,j}$ should be the same for asymptotic linear and strictly linear amplitudes except for $j=0$. Indeed, in (\ref{eq: asymptotic_formula_kappa_2_correction}), (\ref{eq: asymptotic_formula_kappa_2_correction1}), and (\ref{eq: asymptotic_formula_kappa_2_correction2}) we found the asymptotic formula
\begin{equation*}
    \beta_{n,j}\sim \mathcal{B}_{n,j}^{(0)}(\kappa_1,\delta) +\kappa_2\mathcal{B}_{n,j}^{(1)}(\kappa_1,\delta),
\end{equation*}
where
\begin{align*}
    &\mathcal{B}_{n,j}^{(0)}(\kappa_1,\delta) \sim 2^j j!\bigg[\delta_{0,j}\frac{n^{-\delta}}{\Gamma(1-\delta)}-(-1)^{j+n}\frac{n^{-1+2\delta+2\kappa_1} (\log n)^{-1}}{\Gamma(1+2\delta+2\kappa_1)}+\frac{n^{-1-\delta-\kappa_1}(\log n)^{-1}}{\Gamma(1-\delta-\kappa_1)}\bigg]\,,\\
    &\mathcal{B}_{n,.j}^{(1)}(\kappa_1,\delta) \sim -2^j j!\frac{n^{-1-\delta}(\log n)^2}{\Gamma(1-\delta)}\,.
\end{align*}
From these ones can easily see that the first term of $\mathcal{B}_{n,j}^{(0)}(\kappa_1,\delta)$ only contributes to the scalar sector and furthermore,  as $\kappa_2$ is set to zero, the non-scalar partial wave coefficients match precisely with the strict ones given in (\ref{a-deformed_asymptotic_linear_massgap_bound}).

 \begin{figure}[h!]
    \centering
    \includegraphics[width=0.8\linewidth]{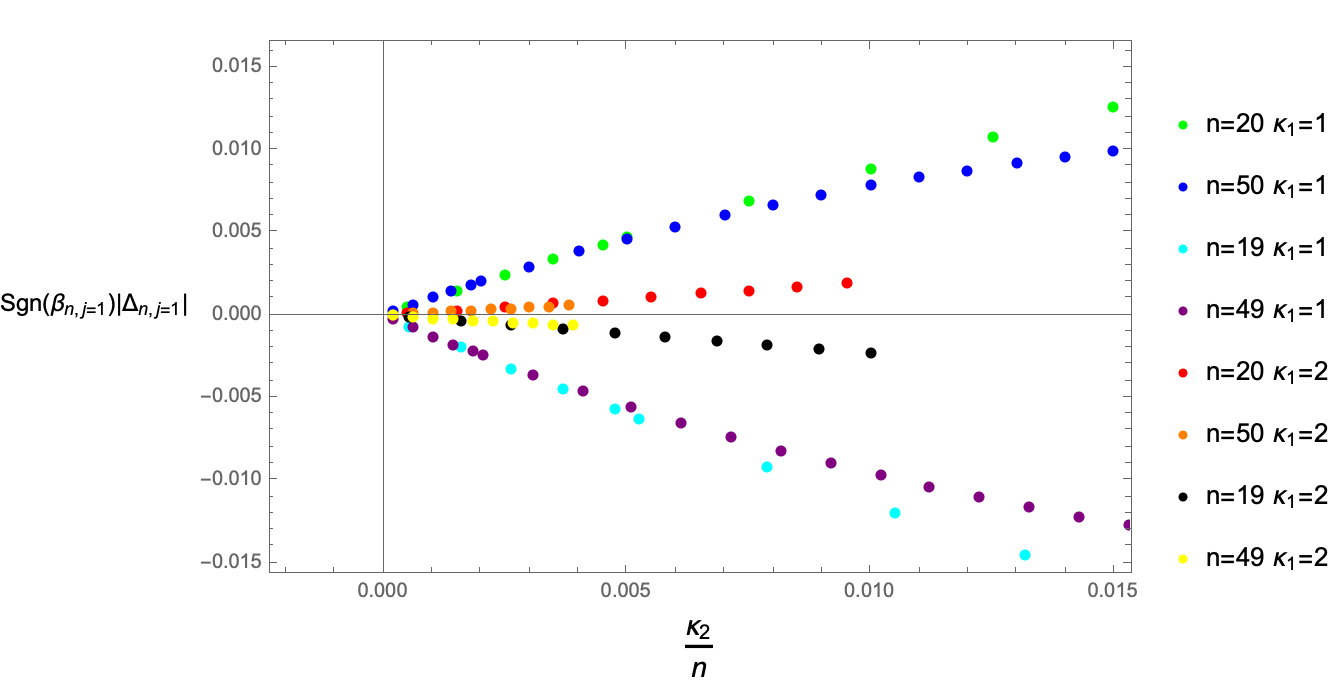}
    \caption{The difference between $\beta_{n,j}(\kappa_2\ne 0)$ and $\beta_{n,j}(\kappa_2=0)$ decreases as the ratio $\kappa_2/n$ decreases. This implies that $\beta_{n,j}(\kappa_2)$ will approach $\beta_{n,j}(\kappa_2=0)$ as $n$ increases. Furthermore, $\beta_{n,j}(\kappa_2)$ preserves the sign-flipping pattern of the linear spectrum coefficients. In this plot, we fixed $(j,\delta,\delta_0)=(1,0,0)$.}\label{deltafig}
\end{figure}

Next, we provide numerical evidence that partial wave coefficients of asymptotic linear amplitudes approach those of linear amplitudes in the large $n$ limit. The partial wave coefficients are given by
\begin{equation}
    B_{n,j}=J_n\alpha_{n,\delta_0}n_{n,\delta_0}\int_{-1}^1dx(1-x)^{j+\delta_0}\partial_x^j\oint_{u=0}\frac{du}{2\pi i}\frac{e^{u (n+1)}}{(e^u-1)^n}\sum_{\alpha}e^{-\nu_\alpha(t)}=J_n\alpha_{n,\delta_0}n_{n,\delta_0}\beta_{n,j}\,.
\end{equation}
We now ignore the prefactors and compare $\beta_{n,j}$ at various $\{\kappa_i\}$. For simplicity, we consider the first correction to the strictly linear spectrum; that is 
\begin{equation}
    f(\mu,\nu)=(\nu+\delta)^2+\kappa_1(\nu+\delta)+\kappa_2-\mu(\nu+\delta)\,.
\end{equation}
We use the measure $\Delta_{n,j}=\big|1-\frac{\beta_{n,j}(\kappa_2)}{\beta_{n,j}(\kappa_2=0)}\big|$ to quantify the difference between the linear spectrum and the $\kappa_2\ne 0$ spectrum.  Figure~\ref{deltafig} shows that the sign of $\beta$ is consistent as $\kappa_2$ is deformed away from 0; that is, the sign flipping phenomenon of partial wave coefficients is preserved. Moreover, in the region $\frac{\kappa_2}{n}\ll1$ there is a universal linear behavior of $\beta_{n,j}(\kappa_2)$ regardless the value of $\kappa_2$. This numerical evidence supports the claim that the partial wave coefficients asymptotically approach those of the linear amplitude in the large $n$ limit.

\subsection{Numerical checks for \eqref{eq: asymptotic_formula_kappa_2_correction}}\label{app:numerical_checks-asympototic}

While performing a numerical check of the asymptotic formula, we found that for a vast majority of the points in $(\delta,\kappa_1)$ space, the formula matches the order and the sign of the exact result of $\beta_{n,j}$. In our numerical analysis, we defined convergence in the following way: (i) The sign of the asymptotic formula matches with the exact result, and (ii) the ratio between the asymptotic formula and exact integral is of order one and approaches $1$ as $n$ increases. The non-convergent points are those where $\delta$ and $\kappa_1$ are small.

\begin{figure}[h!]
    \centering
\includegraphics[width=0.8\linewidth]{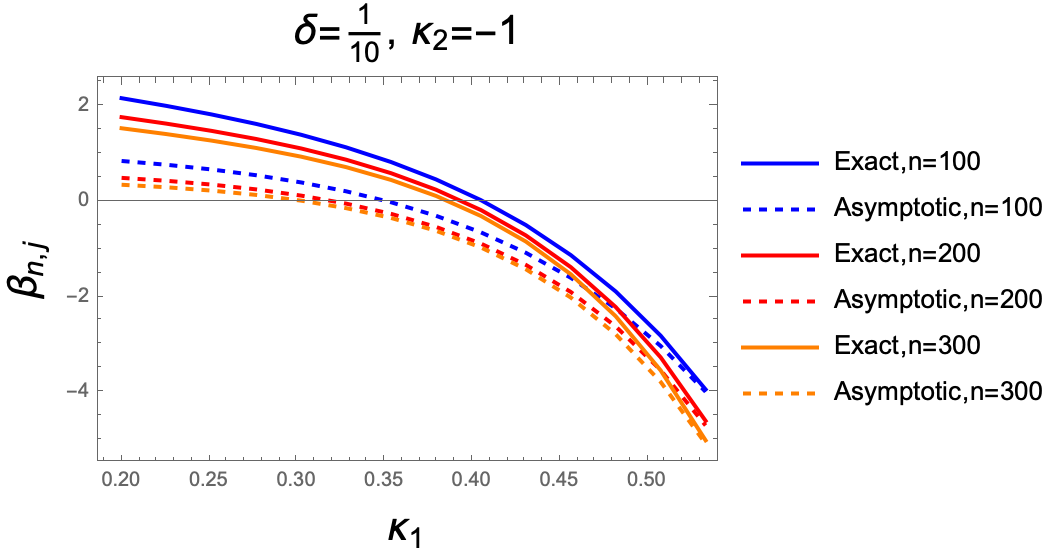}
    \caption{A comparison of asymptotic formula and exact integral at fixed $(\delta,\kappa_2)=(\frac{1}{10},-1)$. The plot shows that the asymptotic formula shows good agreement with the exact formula at $\kappa_1\gtrsim0.5$. We expect the discrepancy at smaller $\kappa_1$ value will vanish as $n$ is taken to larger values.}
    \label{fig:asymptotic_off_sign}
\end{figure}

In order to understand the non-convergence at small $\delta$ and $\kappa_1$, we plot out the values of $\beta_{n,j}$ given by asymptotic formula and the exact integral while fixing $(\delta,\kappa_2)=(\frac{1}{10},-1)$ in Figure~\ref{fig:asymptotic_off_sign}. We noticed that the formula agrees with the exact result well as $\kappa_1\gtrsim 0.5$ and there is a region where the sign of the asymptotic formula and the exact integral differs. As $n$ increases, the curves of the exact coefficient are moving toward negative, and so are the curves of the asymptotic formula. We believed that the discrepancy between the asymptotic formula and the exact result at small $\delta$ and $\kappa_1$ would close after $n$ is taken to be larger values, and the disagreement of the sign is just a consequence that the asymptotic formula needs to go to larger values of $n$ for a good agreement.

\section{Mathematical addendum}\label{app:mathematical_addendum_special_functions}
\subsection{Modified Bessel function $I_{\alpha}(x)$}

The key references for this section are \cite{238f62ea-f226-315b-a579-8f3544634def,watson1995treatise} unless stated otherwise. Modified Bessel functions, denoted as $I_{\alpha}(x)$, are analytic continuations of the usual Bessel function $J_{\alpha}(x)$ into the entire complex plane $x \in \mathbb{C}$. This is often achieved via a simple Wick rotation into the imaginary axis $I_{\alpha}(x) = i^{-\alpha} J_{\alpha}(ix)$, and as a result, these have the series definition 
\begin{align}
I_{\alpha}(x) &=  \sum_{m=0}^\infty \frac{1}{m!\, \Gamma(m+\alpha+1)}\left(\frac{x}{2}\right)^{2m+\alpha}, 
\end{align}
valid for any $\alpha \in \mathbb{C}$, and convergent for all $x$ courtesy to the factorial in the denominator of the summand. These are one of the two independent solutions to the following second-order ODE: 
\begin{equation}
    x^2 \frac{d^2 y}{dx^2} + x \frac{dy}{dx} - \left(x^2 + \alpha^2 \right)y = 0\,,
\end{equation}
valid for any $\alpha \in \mathbb{R}$. For $\alpha=n\in \mathbb{N}$, it can be shown that there exists a particularly useful contour representation of these functions (cf.~\cite{Arfken1985}):
\begin{equation}
    I_{n}(x) = \oint_{t=0}\frac{dt}{2\pi i}\frac{e^{\frac{x}{2}(t+\frac{1}{t})}}{t^{n+1}}\,.
\end{equation}
However, in general, there are no canonical contour representations of these functions. For instance, if $\alpha \in \mathbb{C}$ then the integral representation is given as (see for example \cite{238f62ea-f226-315b-a579-8f3544634def})
\begin{equation}
    I_{\alpha}(x) =\frac{2^{-\alpha}x^{\alpha}}{\sqrt{\pi}\Gamma(\alpha+\frac{1}{2})}\int_{-1}^1dt~(1-t^2)^{\alpha-\frac{1}{2}}e^{tx}~.
\end{equation}
Finally, the modified Bessel functions have the asymptotic expansion as $x\to \infty$ 
\begin{equation}
    I_{\alpha}(x) \sim \frac{e^x}{(2\pi x)^{\frac{1}{2}}}\sum_{k=0}^{\infty}\frac{(\frac{1}{2}-\alpha)_k(\frac{1}{2}+\alpha)_k}{2^kk!x^k}
\end{equation}
for $\alpha \in \mathbb{C}$, which is particularly handy in our calculation of \eqref{eq: asymptotic_formula_kappa_2_correction} in Appendix \ref{app: post_regge_integral_detail}. 

\subsection{Generalized Bernoulli polynomials $B_{\nu}^{(n)}(z)$}

The key references for this section are
\cite{Temme:1996,LOPEZ2010197} unless stated otherwise. The generalized Bernoulli polynomials are generalizations of the Bernoulli numbers, the latter show up in the convergence formulas for arithmetic sums of various positive powers of natural numbers (cf.~\cite{Gradshteyn}). These polynomials are most commonly defined in terms of their generating functions:
\begin{equation}
    \left(\frac{z}{e^z-1}\right)^{n} e^{tz} = \sum_{\nu=0}^{\infty}\frac{B^{(n)}_{\nu}(t)}{\nu!}z^{\nu}\label{eq:bernoulli_sum_formula}\,,
\end{equation}
where $t,z,n\in \mathbb{C}$.
They have the contour representation
\begin{equation}
    B_{\nu}^{(n)}(z) = \Gamma(\nu)\int_{\mathcal{C}}\frac{dt}{2\pi i}\left(\frac{t}{e^{t}-1}\right)^n\frac{e^{tz}}{t^{\nu+1}}\,,
\end{equation}
where $\mathcal{C}$ is a contour that closes around the pole $t=0$ for $\nu\in \mathbb{N}$ or can be written as a Hankel-like contour for more general $\nu \in \mathbb{C}$. This provides an appropriate analytic continuation of \eqref{eq:bernoulli_sum_formula} for $\nu \in \mathbb{C}$. It also turns out to be extremely useful for deriving various asymptotic formulae for the Bernoulli polynomials by deforming $\mathcal{C}$ along the saddle point of the integrand \cite{Temme:1996}.
For $\nu \to \infty$, while $n$ and $z$ are any arbitrary complex numbers, the asymptotic expansion for the generalized Bernoulli polynomials is given as
\begin{equation}
    B_{\nu}^{(n)}(z) \sim \frac{2\nu!\nu^{n-1}}{(2\pi)^{\nu}\Gamma(n)}\left[\cos(\pi\left(2z+n-\frac{1}{2}\nu\right))+\mathcal{O}(\nu^{-1})\right].
\end{equation}
On the other hand for $z\to \infty$, with $n~\text{and}~\nu$ kept fixed, the asymptotic expansion is
\begin{equation}
    B_{\nu}^{(n)}(z) \sim \sum_{k=0}B_{k}^{(n)}\frac{\Gamma(\nu)}{\Gamma(\nu-k)}\frac{z^{n-k}}{k!}~.
\end{equation}
And finally, for $n\to \infty$, while $\rho$ and $z$ are arbitrary complex numbers, the following turns out to be the asymptotic expansion 
\begin{equation}
    \frac{B^{(n+\rho+1)}_{n}(z)}{n!}\sim (-1)^{n}\frac{(\log n)^{\rho}}{n^z}\left[\sum_{s=0}^{k-1}\frac{\Gamma(\rho)}{\Gamma(\rho-s)}\frac{(-1)^s}{(\log n)^{s}}\frac{A_s(z)}{s!}+\mathcal{O}((\log n)^{-k})\right]
\end{equation}
where $k$ is a parameter encoding the precision of the asymptotic expansion, and we have further defined 
\begin{equation}
    A_s(z) \equiv \frac{d^s}{dz^s}\frac{1}{\Gamma(1-z)}\,,
\end{equation}
which can be written in terms of various linear combinations of the polygamma functions $\psi^{(m)}(z)$ for $m\leq s-1$. This particular form of the asymptotic expansion in large $n$ is quite handy in the calculations in section \ref{sec:post_linear_regge_unitarity} and then later in Appendix \ref{app: post_regge_integral_detail}.

\subsection{Incomplete gamma function $\gamma(a;z)$}

The key reference for this section is \cite{NIST:DLMF}. The incomplete gamma function is defined as 
\begin{equation}
    \gamma(a;z) = \int_{0}^zt^{a-1}e^{-t}~dt
\end{equation}
for $\text{Re}(a)>0$ and $z\in \mathbb{C}$. In the limit $z\to \infty$, one recovers the usual gamma function $\Gamma(a)$. The incomplete gamma function inherits its pole structure from its complete cousin and thus has poles in $a$ at non-positive integers. Whenever $a\notin \mathbb{Z}$, $\gamma(a;z)$ has a branch cut in $z$ running between zero and infinity with a phase $e^{2\pi i a}$, and thus
\begin{equation}
    \gamma(a;e^{2\pi i m}z) = e^{2\pi i m a}\gamma(a;z)~~~\text{for}~m\in \mathbb{Z}\,.
\end{equation}
It also satisfies the differential equation
\begin{equation}
    z\frac{d^2w}{dz^2}+(z+1-a)\frac{dw}{dz}=0\,.
\end{equation}
Furthermore, it has a particularly elegant indefinite integral identity
\begin{equation}
     \int x^{b-1}\gamma (a,x)dx=\frac {1}{b}\left(x^{b}\gamma (a,x)-\gamma (a+b,x)\right)\,,\label{eq:indefinite_integral_id_incomp_gamma}
\end{equation}
for $b\in \mathbb{C}$. It has the power series expansion 
\begin{equation}
 \gamma (a,z)=\sum _{k=0}^{\infty }\frac {z^{a}e^{-z}z^{k}}{a(a+1)\dots (a+k)}=z^{a}e^{-z}\sum _{k=0}^{\infty }{\dfrac {z^{k}}{a^{\overline {k+1}}}}\,,
\end{equation}
which comes in handy when deriving the identity \eqref{eq:action_of_derivative}. Finally, it has the following asymptotic expansion as $|z|\to \infty$ with $|\text{arg}~z|<\frac{3\pi}{2}$:
\begin{equation}
     \gamma (a,z)\sim -z^{a-1}e^{-z}\sum _{k=1}^{N}{\frac {\Gamma (s)}{\Gamma (s-k)}}z^{-k}~~~\text{for}~N\geq 1\,.\label{eq:asymptotic_incomp_gamma}
\end{equation}
Equations \eqref{eq:indefinite_integral_id_incomp_gamma} and \eqref{eq:asymptotic_incomp_gamma} are vital in evaluating the asymptotic values of some of the integrals in Appendix \ref{app: post_regge_integral_detail}.

\bibliography{main}
\bibliographystyle{JHEP}

\end{document}